\documentclass[sigconf]{acmart}

\setcopyright{none}
\renewcommand\footnotetextcopyrightpermission[1]{}
\AtBeginDocument{%
  }
    \acmConference{}{}{}
\acmBooktitle{}
\usepackage{tikz}
\usepackage{enumitem}
\usepackage{tabularx}

\usepackage{booktabs, tabularx, array, xcolor, ragged2e, makecell}
\usepackage{enumitem} 

\newcolumntype{Y}{>{\RaggedRight\arraybackslash}X}
\definecolor{thead}{RGB}{245,245,245}
\definecolor{rowalt}{RGB}{250,251,253}
\newcommand{\qcell}[1]{\textit{\color{black!70}``#1''}}

\usepackage{adjustbox}
\usepackage{tikz}
\usetikzlibrary{arrows.meta}
\usepackage{booktabs, tabularx, array, ragged2e, makecell}
\usepackage[table]{xcolor}

\usepackage[skip=0pt plus1pt, indent=15pt]{parskip}

\usepackage[edges]{forest} 
\forestset{
  annotree/.style={
    for tree={
      grow'=east,               
      parent anchor=east,
      child anchor=west,
      anchor=west,
      align=left,
      edge={line width=0.6pt},  
      s sep=6pt,                
      l sep=10pt,               
      font=\small
    },
    before typesetting nodes={
      if n=1{ 
        tikz+={
          \node[font=\bfseries] at (!.child anchor) {};
        }
      }{}
    }
  }
}
\begin{document}

\author{Hina Saeeda}
\email{hinasa@chalmers.se}

\authornotemark[1]
\affiliation{%
  \institution{Chalmers University of Technology and University of Gothenburg}
  \country{Sweden}
}

\author{Tommy Johansson }
\affiliation{%
  \institution{Kognic AB }
  \city{Gothenburg}
  \country{Sweden}
}
\email{tommy.johansson@kognic.com}

\author{Mazen Mohamad}
\affiliation{%
  \institution{RISE Research Institutes}
  \city{Gothenburg}
  \country{Sweden}}
\email{mazen.mohamad@ri.se}

\author{Eric Knauss}
\affiliation{%
  \institution{Chalmers University of Technology and University of Gothenburg}
  \city{Gothenburg}
  \country{Sweden}
}
\email{eric.knauss@cse.gu.se}

\title{Data Annotation Quality Problems in AI-Enabled Perception System Development} 

\renewcommand{\shortauthors}{ Saeeda et al.}

\begin{abstract}
Data annotation is a critical yet error-prone activity in developing AI-enabled perception systems (AIePS) for automated driving.
Annotation quality directly affects AI model development performance, safety and reliability.
Yet, there is little empirically grounded understanding of how annotation errors arise and propagate across the multi-organisational automotive supply chain.
This study investigates the types, causes, and effects of data annotation errors through a multi-organisation case study involving six companies (\textit{e.g.,} technology and AI companies that develop software for advanced driver assistance systems and self-driving cars) and four research institutes in Europe and the UK. We conducted $19$ semi-structured interviews with $20$ experts ($\approx50$ hours of transcripts) and applied a six-phase thematic analysis.
The resulting data annotation errors taxonomy identifies $18$ recurring error types across three major data-quality dimensions: \textit{completeness} ( attribute omission, missing feedback loop, privacy/compliance omission, edge-case omission, selection bias, sensor synchronisation issues), \textit{accuracy} (wrong class label, bounding-box errors, granularity mismatch, insufficient guidance, bias-driven errors), and \textit{consistency} (inter-annotator disagreement, ambiguous instructions, lack of purpose knowledge, misaligned hand-offs, limited review and logging, lack of frameworks and standards, cross-modality misalignment). For research rigour and triangulation, we further validated the proposed taxonomy of data annotation errors with industry. Practitioners confirmed the usefulness of the taxonomy for root-cause analysis of data annotation errors, supplier quality reviews, new project onboarding, and optimising data annotation guidelines. Practitioners described it as a “failure-mode catalogue” comparable to FMEA (Failure Mode and Effects Analysis).
By framing annotation quality as an AI-enabled system development lifecycle and supply-chain concern, this work advances SE4AI by providing a shared vocabulary, diagnostic checklist, and actionable guidance for trustworthy AIePS development.
\end{abstract}





\maketitle
\section{Introduction}

AI-enabled automotive perception systems (AIePS) are central to automated driving, supporting object detection, tracking, and classification for enhanced safety and efficiency~\cite{bachute2021autonomous}. These systems underpin advanced driver assistance systems (ADAS), which enhance driving safety, cost efficiency, and convenience~\cite{kukkala2018advanced, khattak2021taxonomy}. Core perception capabilities such as \emph{pedestrian detection}, \emph{traffic sign recognition}, and \emph{obstacle avoidance} serve as essential inputs to ADAS functions by interpreting sensor data from cameras, radar, LiDAR, and ultrasonics~\cite{chacon2015detecting, heyn2023automotive, sharma2020evaluation}. Within the AIePS development lifecycle, data annotation, the process of labelling raw sensor data and converting it into structured datasets, remains one of the most expensive and error-prone tasks. Ultimately, the performance of these systems depends on the quality of annotated data used for training and validation~\cite{najafi2024performance, heyn2023automotive}.

\vspace{0.5em}
\noindent
\textbf{Motivation.} While annotation quality is widely recognised as critical for AIePS, the systematic impact of data annotation errors on system performance remains insufficiently understood. In safety-critical domains such as autonomous driving, even minor annotation errors such as missed or inconsistently labelled objects can cascade into unsafe behaviours and degraded model reliability~\cite{chen2022road, yang2023uncertainties, galvao2023pedestrian, zhong2022detecting}.  However, how such errors originate, propagate across the AIePS development lifecycle, and ultimately impact system performance remains underexplored.~\cite{dey2023multi, samuktha2024framework, mohammedali2023influence}.  This gap is amplified in multi-organisational automotive supply chains, where heterogeneous tools, inconsistent annotation practices, and fragmented quality assurance processes increase error propagation. Addressing this issue is essential to shift from reactive correction toward proactive, standardised, and trustworthy annotation processes~\cite{fredriksson2020data}. 

\textbf{Research Gap.} Prior work has focused on isolated error types or technical solutions such as weak supervision and automated labelling~\cite{peters2024generalizable,penquitt2025label,schubert2024identifying}.
Crucially, there are no comprehensive, empirically grounded studies that systematically capture the full spectrum of annotation errors across multi-organisational automotive supply chain contexts. In this domain, different tiers produce annotated datasets, develop AIePS, test them, and deliver integrated products to original equipment manufacturers (OEMs). The lack of an industry validated, comprehensive knowledge base (e.g a detailed taxonomy)  of annotation errors hinders alignment of processes, knowledge sharing, and quality assurance across organisational boundaries.

\textbf{Study Aim.} Annotation quality is not merely a data issue but an AIePS  developemnt lifecycle concern, as errors propagate from annotated data through the AI model to final system decisions. Accordingly, this study aims to provide insights and evaluations of the data annotation errors to support proactive, lifecycle-oriented quality assurance. This study provides a taxonomy of data annotation errors and a classification of their root causes and impacts. This study also provides insights into how  companies can use this information to their advantage after validating the usefulness of the proposed taxonomy with industrial practitioners.

\vspace{0.5em}
\noindent
\textbf{Research Questions.} Guided by these objectives, this multiorganisational case study investigates:
\begin{itemize}
  \item \textbf{RQ1:} What are the different types of data annotation errors, their causes, and how do they affect AIePS development and performance?
  \item \textbf{RQ2:} How do practitioners perceive the usefulness of the proposed taxonomy of data annotation errors?
\end{itemize}

\vspace{0.5em}
\noindent
\textbf{Methodology Overview.} We conducted 19 semi-structured interviews with $20$ participants from six international companies and four research institutes. Using multi-level triangulation, integrating case study data, thematic analysis, and expert validation, the study ensures both empirical depth and practical relevance. 

\vspace{0.5em}
\noindent
\textbf{Novelty.} Our study is novel in (a) it presents the first systematic multi-organisational empirical study in the automotive domain covering the full supply chain, (b) it introduces an industry validated taxonomy of data annotation errors with practical relevance, and (c) it bridges data quality assurance (QA) with SE4AI practice by framing annotation errors as AIePS development lifecycle bottlenecks rather than isolated data issues.

\section{Background and Related Work}
\label{sec:Background}


\textbf{The AIePS developemnt Lifecycle and  Role of Data Annotation.}
The development of AI-enabled systems involves multiple stages, data collection, annotation, model training, evaluation, deployment, and monitoring~\cite{demrozi2021towards, najafi2024performance}. Among these, data annotation—the transformation of raw sensor data into labelled datasets—is a crucial and resource-intensive phase~\cite{fredriksson2020data}. By labelling images, point clouds, or sensor sequences with bounding boxes, segmentation masks, or object classes, annotators create the ground truth essential for model learning, where annotation quality directly impacts performance, generalisation, and decision accuracy in safety-critical domains~\cite{heyn2023automotive, galvao2023pedestrian}.

\textbf{Data Annotation Error.} The annotation errors are incorrect, incomplete, or inconsistent labels in training or validation data~\cite{kruger2022keynote, klie2024analyzing}, arising from human variability, ambiguous guidelines, limited tools, or weak quality assurance~\cite{habibullah2023requirements, mohammedali2023influence}. Examples include misclassifying pedestrians on scooters, inconsistently labelling occlusions, or omitting small objects, all of which introduce noise that weakens model learning~\cite{samuktha2024framework, yang2023uncertainties}. These errors can propagate throughout the AI lifecycle, leading to false detections, sensor misinterpretations, and reduced safety in real-world driving~\cite{chen2022road, zhong2022detecting}.

\textbf{Data Quality Dimensions.}
 Research identifies three core dimensions of data quality: completeness, accuracy, and consistency as key indicators of reliable data representation~\cite{wand1996anchoring,pipino2002data,batini2009methodologies,beck2023quality}. Originally established in information systems~\cite{wand1996anchoring,pipino2002data}, this framework now underpins modern AI models and annotation quality assessment. Our taxonomy builds upon these classical dimensions. Completeness reflects whether all required data are present and relevant real-world entities or attributes are captured~\cite{batini2009methodologies,beck2023quality}. In annotation, missing objects or scenarios (\textit{e.g.,} pedestrians in poor lighting) represent completeness errors that reduce dataset representativeness.
Accuracy measures how closely annotations (\textit{e.g., }bounding boxes, segmentation masks) match true real-world conditions~\cite{pipino2002data,batini2009methodologies}; misaligned boxes or incorrect labels exemplify accuracy errors.
Consistency concerns uniformity across annotators, scenes, or systems~\cite{batini2009methodologies,beck2023quality}. Inconsistent labelling of similar objects or frames constitutes a consistency error that undermines reproducibility, fairness, and reliability.

\textbf{Annotation Quality Across the AI Lifecycle in Industrial Supply Chains.}
AIePS development spans a complex supply chain of OEMs, Tier-$1$, and Tier-$2$ suppliers~\cite{heyn2023automotive, fredriksson2020data}. Each uses distinct annotation tools, standards, and quality controls, resulting in heterogeneous datasets where early-stage errors can propagate through model training, validation, and integration~\cite{mohammedali2023influence}. Consequently, annotation quality represents a dynamic, lifecycle-spanning assurance challenge that directly influences the dependability, maintainability, and trustworthiness of AI-enabled systems~\cite{dey2023multi, samuktha2024framework}.

\section{Research Methodology}
\label{methodology}

\begin{figure}[htbp]  
    \centering
    \includegraphics[width=0.5\textwidth]{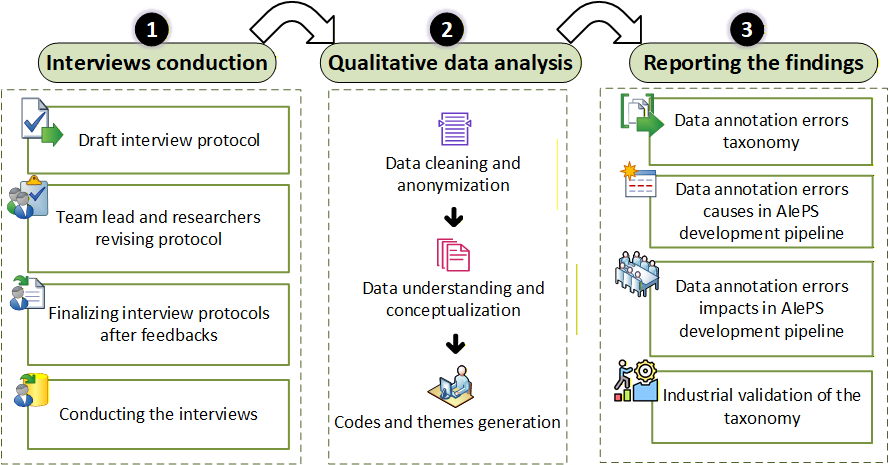} 
    \caption{ Research method followed. }
   \label{fig:method}
\end{figure}

We followed the qualitative case study guidelines established by Runeson and Höst et al.~\cite{runeson2009guidelines}, and aligned our study design and reporting with the empirical standards proposed by Ralph et al.~\cite{ralph2020empirical} to ensure transparency, triangulation, and validity (see Figure~\ref{fig:method}).


\textbf{Research Context} This study is part of Project X, which aims to enhance academia-industry collaboration by developing concepts, models, and techniques for effective and safe AIePS productions.
We conducted $19$ semi-structured interviews~\cite{runeson2009guidelines} with $20$ participants from industry and academia.
These interviews were performed online between October $2024$ and April $2025$, each lasting $\geq 120$ minutes.
ID16 participated in two separate interviews (labelled 16A and 16B), each lasting approximately four hours. In contrast, participants ID18 and ID19 took part in a single focus group session, counted as one interview (see Table \ref{tab:expert_table}). In total, this amounts to over \textbf{\textit{50 hours of rich transcripts}}.
 Using purposeful sampling~\cite{ralph2020empirical}, we selected participants based on their roles, organisational context, and experience with data annotation or AI model development in autonomous driving. Interviews were recorded, transcribed, manually corrected, and concluded with participant feedback. This approach ensured diverse and in-depth insights into annotation quality, associated errors, and their impact on the AIePS development life cycle.

\textbf{Companies Context} To capture diverse perspectives, we conducted fourteen interviews with representatives from six companies engaged in AIePS development. We included the entire supply chain: one OEM, three Tier-$1$ suppliers, two Tier-$2$ suppliers, a perception training company, and a code analysis firm, all based in two European countries and the UK. 

Additionally, five interviewees from four research institutes, including a state-owned European institute and leading universities from Europe and the UK, helped bridge theory and practice across AI, data annotation, quality, and regulation. Thematic Saturation was reached with Interview $15$, with no new information or candidate themes emerging in Interviews $16$, $17$, $18$, and $19$.

\textbf{Participants} As shown in Table \ref{tab:expert_table}, targeted experts included leading contributors to AIePS, such as safety standard developers, founders of relevant companies, and the chief investigator of top-tier research labs, ensuring deep domain expertise. This is a niche domain where only a few globally recognised organisations and experts are active. We deliberately reached out to these leading actors and succeeded in including the most relevant and established experts in this emerging field.

\begin{table}[!ht]
\vspace{-0.3cm}
\centering
\scriptsize
\caption{Categorisation of interviewees based on company type and specialisation, including their roles and years of experience}
    \label{tab:expert_table}
\begin{tabularx}{\linewidth}{llll}
\toprule

        \textbf{Company ID: Type} & \textbf{Focus Area} & \textbf{Expert (Years of Experience)} \\
        \midrule
        A: Tier 2 & Data Annotation & ID1: Annotation Expert (6) \\
                            && ID2: Perception Expert (10) \\
                            && ID3: Quality  Expert (9) \\
                            \midrule
        B: Tier 1 & Safety Software & ID4: Machine Learning Expert (7) \\
                             &&      ID5: Data Scientist (11) \\
                             \midrule
         C: University & Research & ID6: Senior Researcher (9) \\
         \midrule
         D: University & Research & ID7: Researcher (8) \\
         \midrule
         E: University & Research & ID8: Researcher (5) \\
         \midrule
          F: Research Institute & Research & ID9: Researcher (10)\\
                                     &&   ID10: Researcher (20) \\
                                     \midrule
        G: OEM & Automotive & ID11: Machine Learning Expert (5) \\
                           && ID12: V\&V Expert (5) \\
                          &&  ID13: Data Engineer (4) \\
                          &&  ID14: Researcher (3) \\
                          \midrule
         H: Tier 1 & Safety Software & ID15: Researcher (10) \\
         \midrule
         I: Tier 2 & Quality Assurance & ID16 A: Quality  Expert (18) \\
         I: Tier 2 & Quality Assurance & ID16 B: Quality  
         Expert (18) \\
         \midrule
         J: Tier 1 & Digital Solution & ID17: Head of Research (17)\\
                                  &&   ID18,19: Research Engineer (1,2)   \\

        \bottomrule
    \end{tabularx}
    
\end{table}

\textbf{Interview Preparation:} A semi-structured interview guide~\cite{ralph2020empirical} was developed iteratively by four researchers in collaboration with three industrial experts experienced in data annotation and AIePS, thereby establishing content validity. The guide featured open-ended, neutral, and clear questions aligned with the research questions and relevant literature presented in the background and related work section~\ref{sec:Background}. To enhance clarity, structure, and duration, the guide was piloted with two independent industry partners (not part of the final study), representing annotation requirements (Tier~$1$) and production (Tier~$2$), whose feedback informed refinements.  (For details, see: \href{https://dataverse.harvard.edu/previewurl.xhtml?token=da9478c0-7653-48cb-bc5f-b8ae1d0810e8}{\textcolor{blue}{→ Interview Guide}})

\textbf{Data Analysis:} 
We applied a six-phase thematic analysis process following Braun and Clarke \textit{et al.}~\cite{braun2006using}. 
The analysis was guided by a combined deductive–inductive codebook, and the units of analysis were $19$ interview transcripts, each coded in text chunks of approximately three sentences. 
Across all transcripts, about $1050$ text segments (coded units) were analysed.

\textbf{(1) Familiarisation:} 
Two researchers transcribed, anonymised, and validated all interviews to ensure ethical integrity and data accuracy. 
Through multiple readings, they familiarised themselves with the data, developing an initial understanding of emerging patterns and potential themes. 

\textbf{(2) Initial Coding:} 
A deductive codebook was first developed based on the research questions and interview guide
(see:\href{https://dataverse.harvard.edu/previewurl.xhtml?token=da9478c0-7653-48cb-bc5f-b8ae1d0810e8}{\textcolor{blue}{→ Codebook}} and 
\href{https://dataverse.harvard.edu/previewurl.xhtml?token=da9478c0-7653-48cb-bc5f-b8ae1d0810e8}{\textcolor{blue}{→ Codebook-Aligned Initial Codes}}). 
During open coding, inductive codes were added as new insights emerged. 
Each transcript (ID1–ID19) was coded systematically, resulting in a comprehensive set of detailed codes extracted from all interviews 
(see: \href{https://dataverse.harvard.edu/previewurl.xhtml?token=da9478c0-7653-48cb-bc5f-b8ae1d0810e8}{\textcolor{blue}{→ Complete List of Emerging Codes from Transcripts}}).

\textbf{(3) Code Validation:} 
Two researchers independently coded all transcripts and cross-verified results, achieving a strong inter-coder reliability (Cohen’s $\kappa = 0.8$) corresponding to  80\% agreement. The remaining minor differences (about 20\%) were resolved through discussion and consensus, ensuring the consistency and credibility of the coding process.

\textbf{(4) Theme Development:} 
All codes were clustered into potential themes, related to the three key data quality dimensions \textit{Completeness, Consistency, and Accuracy} and
aligned with the predefined research questions 
(see: \href{https://dataverse.harvard.edu/previewurl.xhtml?token=da9478c0-7653-48cb-bc5f-b8ae1d0810e8}{\textcolor{blue}{→ Detailed Themes Development}}). 
A consolidated analysis produced an initial \textit{Master List of Data Annotation Errors}, capturing all identified errors across the AIePS development lifecycle. The master list compiles identified errors organised by their nature, frequency, and impact, with representative examples forming the basis for the final taxonomy of annotation errors.

\textbf{(5) Theme Refinement:} 
The five researchers collaboratively reviewed, merged, and refined overlapping codes through a three-stage iterative process to ensure analytical robustness. 
In the first round, all codes within each theme were compiled, and their frequencies (\textit{i.e.,} the number of mentions across interviews were recorded. 
Across the three quality dimensions, we identified 23 codes under \textit{Completeness}, $23$ under \textit{Consistency}, and $18$ under \textit{Accuracy}. 
In the second round, codes mentioned three times or fewer were classified as low-frequency. 
In the final round, these were removed, retaining only the most recurrent data annotation errors consistently mentioned by multiple experts. 
This resulted in six high-frequency codes for \textit{Completeness}, seven for \textit{Consistency}, and five for \textit{Accuracy} (see: \href{https://dataverse.harvard.edu/previewurl.xhtml?token=da9478c0-7653-48cb-bc5f-b8ae1d0810e8}{\textcolor{blue}{→  Final Refine Data Annotation Errors List}}).

\textbf{(6) Reporting:} 
 The final themes and associated codes were verified, documented, and linked with representative quotes to maintain transparency and traceability.  Five researchers collaboratively reviewed the complete analysis using Excel, OneDrive, and Miro, ensuring structured documentation and analytic rigour. 

\textbf{Replication Package:}
We make the interview guide, codebook, thematic analysis themes, and validation survey available to support future researchers in replicating the study. The replication package is available on HARVARD Dataverse \footnote{\href{https://dataverse.harvard.edu/previewurl.xhtml?token=da9478c0-7653-48cb-bc5f-b8ae1d0810e8}{\textbf{The replication data associated with this study is publicly accessible and available online here.}}}.

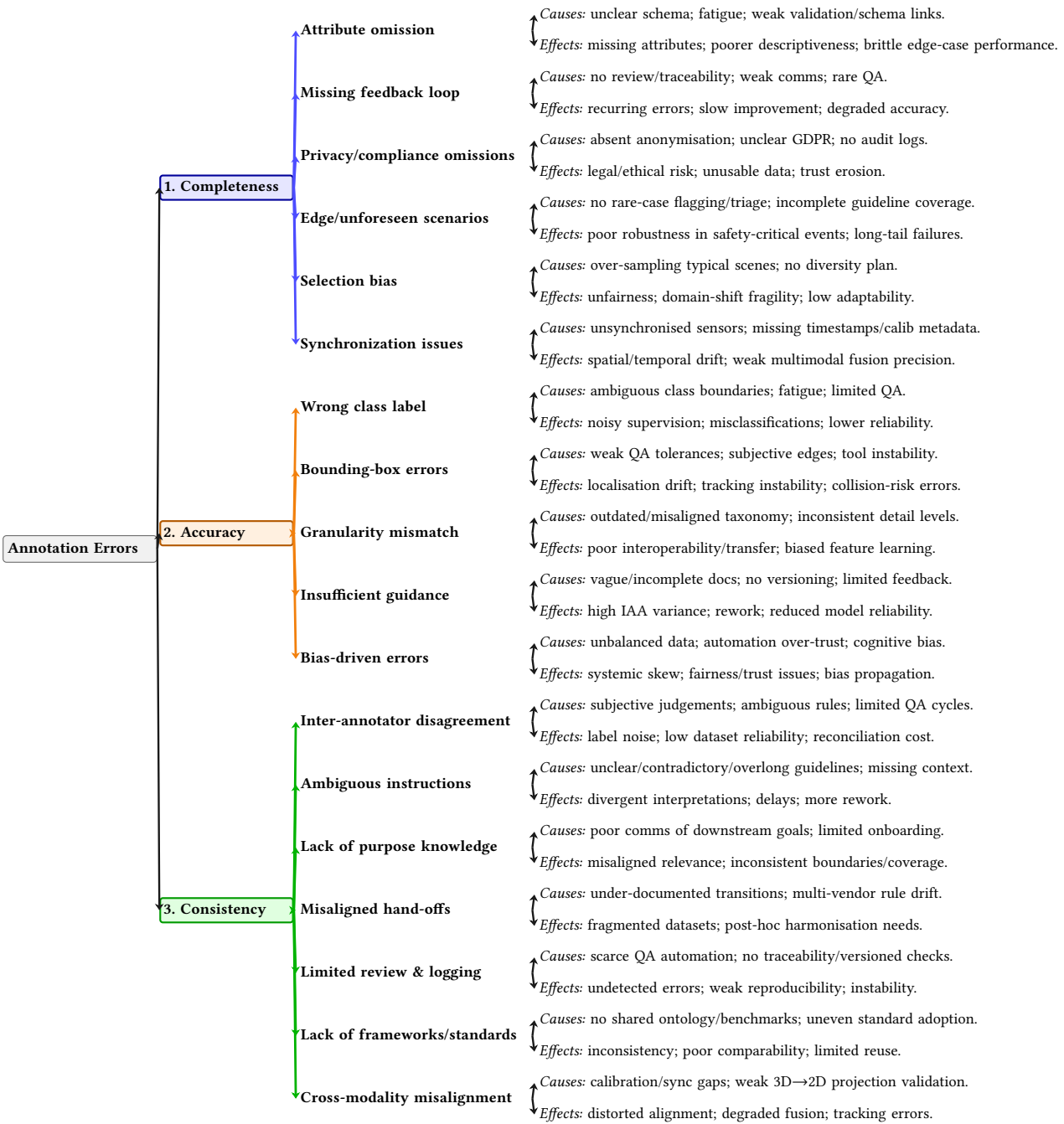
\begin{figure*}[t]
\raggedright
\hspace*{-0.1cm} 
\begin{adjustbox}{max width=\textwidth}
\begin{forest}
for tree={
  grow'=east,
  parent anchor=east,
  child anchor=west,
  anchor=west,
  align=left,
  edge={draw=black!90, line width=0.9pt, -{Stealth[length=3pt,width=4pt]}},
  s sep=1.8pt,       
  l sep=0.9pt,       
  font=\footnotesize,  
  rounded corners=1.2pt,
  inner sep=1.5pt,
  minimum height=0pt
},
where level=1{}{text width=auto, inner sep=2pt},
where level=2{}{text width=3.8cm},   
where level>=3{}{text width=3.6cm},  
[{\textbf{Annotation Errors}}, 
  fill=gray!11, 
  draw=black!60,
  rounded corners=1.5pt, 
  inner sep=2pt,
  text width=2.3cm,       
  align=center,
  font=\footnotesize\bfseries,
  [{\textbf{1. Completeness}}, 
    fill=blue!10, 
    draw=blue!60!black, 
    thick, 
    rounded corners=1.2pt, 
    inner sep=1.5pt,
    text width=2.0cm,      
    align=center,
    font=\footnotesize\bfseries,
    for children={edge={draw=blue!70, line width=1.05pt, -{Stealth[length=3pt,width=4pt]}}},
    [\textbf{Attribute omission}
      [\textit{Causes:} unclear schema; fatigue; weak validation/schema links.]
      [\textit{Effects:} missing attributes; poorer descriptiveness; brittle edge-case performance.]
    ]
    [\textbf{Missing feedback loop}
      [\textit{Causes:} no review/traceability; weak comms; rare QA.]
      [\textit{Effects:} recurring errors; slow improvement; degraded accuracy.]
    ]
    [\textbf{Privacy/compliance omissions}
      [\textit{Causes:} absent anonymisation; unclear GDPR; no audit logs.]
      [\textit{Effects:} legal/ethical risk; unusable data; trust erosion.]
    ]
    [\textbf{Edge/unforeseen scenarios}
      [\textit{Causes:} no rare-case flagging/triage; incomplete guideline coverage.]
      [\textit{Effects:} poor robustness in safety-critical events; long-tail failures.]
    ]
    [\textbf{Selection bias}
      [\textit{Causes:} over-sampling typical scenes; no diversity plan.]
      [\textit{Effects:} unfairness; domain-shift fragility; low adaptability.]
    ]
    [\textbf{Synchronization issues}
      [\textit{Causes:} unsynchronised sensors; missing timestamps/calib metadata.]
      [\textit{Effects:} spatial/temporal drift; weak multimodal fusion precision.]
    ]
  ]
  [{\textbf{2. Accuracy}}, 
    fill=orange!12, 
    draw=orange!70!black, 
    thick, 
    rounded corners=1.2pt, 
    inner sep=1.5pt,
    text width=2.0cm,      
    align=center,
    font=\footnotesize\bfseries,
    for children={edge={draw=orange!80!brown, line width=1.05pt, -{Stealth[length=3pt,width=4pt]}}},
    [\textbf{Wrong class label}
      [\textit{Causes:} ambiguous class boundaries; fatigue; limited QA.]
      [\textit{Effects:} noisy supervision; misclassifications; lower reliability.]
    ]
    [\textbf{Bounding-box errors}
      [\textit{Causes:} weak QA tolerances; subjective edges; tool instability.]
      [\textit{Effects:} localisation drift; tracking instability; collision-risk errors.]
    ]
    [\textbf{Granularity mismatch}
      [\textit{Causes:} outdated/misaligned taxonomy; inconsistent detail levels.]
      [\textit{Effects:} poor interoperability/transfer; biased feature learning.]
    ]
    [\textbf{Insufficient guidance}
      [\textit{Causes:} vague/incomplete docs; no versioning; limited feedback.]
      [\textit{Effects:} high IAA variance; rework; reduced model reliability.]
    ]
    [\textbf{Bias-driven errors}
      [\textit{Causes:} unbalanced data; automation over-trust; cognitive bias.]
      [\textit{Effects:} systemic skew; fairness/trust issues; bias propagation.]
    ]
  ]
  [{\textbf{3. Consistency}}, 
    fill=green!12, 
    draw=green!60!black, 
    thick, 
    rounded corners=1.2pt, 
    inner sep=1.5pt,
    text width=2.0cm,      
    align=center,
    font=\footnotesize\bfseries,
    for children={edge={draw=green!70!black, line width=1.05pt, -{Stealth[length=3pt,width=4pt]}}},
    [\textbf{Inter-annotator disagreement}
      [\textit{Causes:} subjective judgements; ambiguous rules; limited QA cycles.]
      [\textit{Effects:} label noise; low dataset reliability; reconciliation cost.]
    ]
    [\textbf{Ambiguous instructions}
      [\textit{Causes:} unclear/contradictory/overlong guidelines; missing context.]
      [\textit{Effects:} divergent interpretations; delays; more rework.]
    ]
    [\textbf{Lack of purpose knowledge}
      [\textit{Causes:} poor comms of downstream goals; limited onboarding.]
      [\textit{Effects:} misaligned relevance; inconsistent boundaries/coverage.]
    ]
    [\textbf{Misaligned hand-offs}
      [\textit{Causes:} under-documented transitions; multi-vendor rule drift.]
      [\textit{Effects:} fragmented datasets; post-hoc harmonisation needs.]
    ]
    [\textbf{Limited review \& logging}
      [\textit{Causes:} scarce QA automation; no traceability/versioned checks.]
      [\textit{Effects:} undetected errors; weak reproducibility; instability.]
    ]
    [\textbf{Lack of frameworks/standards}
      [\textit{Causes:} no shared ontology/benchmarks; uneven standard adoption.]
      [\textit{Effects:} inconsistency; poor comparability; limited reuse.]
    ]
    [\textbf{Cross-modality misalignment}
      [\textit{Causes:} calibration/sync gaps; weak 3D→2D projection validation.]
      [\textit{Effects:} distorted alignment; degraded fusion; tracking errors.]
    ]
  ]
]
\end{forest}
\end{adjustbox}
\caption{Taxonomy of data annotation errors in AI-enabled perception systems, organised into three main categories: Completeness, Accuracy, and Consistency.}
\label{fig:taxonomy}
\end{figure*}

\section{Findings}

The Findings section presents a taxonomy of data annotation errors, highlighting their causes and impacts (Figure~\ref{fig:taxonomy}). It also reports expert validation results confirming the taxonomy’s usefulness, industrial relevance, and practical applicability (Section~\ref{sec:validation}).

\subsection{RQ1: Data Annotation Errors, their Types, Causes and Impacts in AIePS Development}

\textbf{(1) Completeness Errors.} 
This section addresses completeness errors related to both data coverage and the data annotation processes that sustain and verify that coverage over time. 

\textbf{(1.1) Attribute Omission:} Missing or incomplete attributes (\textit{e.g.,} colour, state, occlusion, or diversity
attributes) leave datasets insufficiently descriptive. As one participant noted, \textit{“Another issue is adding proper metadata [attribute]… otherwise, older datasets become outdated or inconsistent.”} (ID10) Causes include unclear requirements, annotator fatigue, and lack of validation or schema linkages (ID5–ID10). Such omissions degrade reliability and performance in edge case scenarios.

\textbf{(1.2) Missing Feedback Loop:} This error refers to the absence of a systematic review and refinement process that allows recurring annotation errors to persist across iterations. As highlighted, \textit{“The most striking aspect missing from this process flow is the feedback loop… teams cannot adjust their approach during annotation, and mistakes get repeated.”} (ID3) Weak communication, rare quality reviews, and tools lacking traceability hinder continuous improvement (ID9–ID18), slowing iterations and degrading perception quality. The “Missing feedback loop” is categorised under completeness because it represents a process through which completeness fails to be achieved or maintained, rather than an isolated data issue.

\textbf{(1.3) Privacy and Compliance Omission:} This error refers to the insufficient enforcement or documentation of privacy and compliance measures during data annotation. Inadequate anonymisation or weak enforcement of GDPR-like policies introduces ethical and legal risks. Participants noted, \textit{“Ensuring data privacy… is critical when dealing with sensitive data,”} (ID15) and \textit{“GDPR requirements remain somewhat unclear and there is an inherent risk in misinterpreting these guidelines”} (ID9). Missing anonymisation tools or audit logs (ID13–ID18) reduce visual uniformity and trustworthiness. Although “Privacy and Compliance Omission” primarily relates to ethical and legal governance, it is included under the completeness theme as it directly influences dataset coverage. Weak enforcement or unclear interpretation of privacy requirements often leads to the omission of sensitive samples or attributes, thereby reducing the representativeness and completeness of annotated data.

\textbf{(1.4) Edge cases /Unforeseen Scenarios Omission:}  This error refers to the omission of rare, unexpected, or safety-critical cases that fall outside predefined annotation guidelines. Rare or unforeseen scenarios are often ignored. One participant explained, \textit{“At the same time, we should acknowledge the existence of unknown scenarios, particularly edge cases. The challenge with unknown scenarios is that we cannot write explicit guidelines for them because they are unknown.”} (ID12) Another added, \textit{“Ambiguities arose—whether a pedestrian on a scooter should be annotated as relevant or not.”} (ID6) Missing mechanisms to flag unusual cases (ID8, ID13) limit robustness in safety-critical settings.

\textbf{(1.5) Selection Bias:} Overrepresentation of common conditions (\textit{e.g.,} daylight, clear weather) creates unbalanced datasets. \textit{“Some projects focus on specific events… this inherently introduces bias.”} (ID5) and \textit{“Most datasets capture sunny conditions… models struggle in rain or at night.”} (ID18) Lack of planning and diversity requirements (ID14–ID17) reduces model adaptability and fairness.

\textbf{(1.6) Synchronisation \textbackslash{} Calibration Issues:} This error refers
to the misalignment or incomplete calibration of multi-sensor systems such as cameras, radar, and LiDAR used during data 
collection and annotation. Misaligned or unsynchronised sensors produce annotation drift. As noted, \textit{“You need to synchronise sensors so that data points align in time and space… otherwise, they show up in different places.”} (ID10) Missing timestamp checks and calibration metadata (ID16–ID18) reduce precision in multi-sensor fusion.

\textbf{(2) Accuracy Errors.} Under this theme, we address both procedural and data related factors contributing to accuracy errors. We are adopting a holistic life cycle perspective rather than a purely data-centric view. 

\textbf{(2.1) Wrong Class Label:} This error refers to the incorrect assignment of semantic categories to objects or scenes, where visually or contextually similar instances are labelled under the wrong class.  Misclassifications occur when annotators confuse similar categories. \textit{“In autonomous driving, you have different classes of vehicles....OK, this is a car, this is a truck, this is a van … different people might annotate the same object differently.”} (ID18) Ambiguous class boundaries, annotator fatigue, and poor validation (ID2–ID7) lead to noisy supervision and degraded model reliability.

\textbf{(2.2) Bounding Box Errors:} This error refers to inaccurate, inconsistent, or corrupted bounding boxes used to localise objects in images or sequences. Bounding-box errors arose from weak quality assurance tolerances, ambiguous guidelines for occlusions, subjective boundary interpretations, annotator fatigue, and tool instability, causing inconsistent rendering or corrupted coordinate data. One participant shared, \textit{“Some clients insist on pixel-perfect accuracy like a 10-pixel margin.”} (ID2) Another added, \textit{“When annotating objects… avoid optimistic annotations; underestimating size causes issues in collision avoidance.”} (ID12) Missing QA tolerances and unstable rendering tools (ID8–ID14) amplify spatial drift and tracking instability.

\textbf{(2.3) Granularity Mismatch:} This error refers to inaccuracies in the level of annotation detail, either overly fine-grained or overly coarse, relative to the intended schema or perception task. Misalignment between annotation detail and schema depth leads to inconsistency. \textit{“If the requirement specifies capturing the entire head but annotators only mark facial features, this introduces bias.”} (ID8) and \textit{“We used a taxonomy… it didn’t include trams… the taxonomy was from the US.”} (ID17). Outdated taxonomies (ID9, ID18) reduce dataset interoperability and model transferability. Inconsistent levels of detail, such as incomplete object coverage or missing 3D dimensions (ID8, ID18), impair model calibration and feature learning, ultimately degrading recognition accuracy and the ability of AIePS to adapt across diverse operational domains.

\textbf{(2.4) Insufficient Guidance:}  This error refers to vague, incomplete, or ambiguous annotation guidelines that fail to provide clear operational instructions for annotators. Unclear and incomplete annotation guidelines lead to subjective interpretation. \textit{“If annotators are given vague guidelines… they may not annotate consistently.”} (ID2) and \textit{“Initially, it wasn’t clear how to annotate in cases where visibility was poor.”} (ID5) Insufficient guidance resulted from incomplete documentation of quality criteria, ambiguous or evolving rules, limited annotator expertise and feedback, and tool limitations such as missing guideline prompts, version control, and traceability to design goals. Ambiguous or incomplete annotation guidelines increase inter-annotator variability, introduce systematic noise, and weaken label consistency, ultimately reducing model accuracy, reliability, and trust in AIePS (ID5, ID6, ID7, ID16). Although “Insufficient Guidance” stems from process or documentation shortcomings, it is classified as an accuracy error because it directly leads to incorrect or suboptimal labelling. Ambiguous or evolving guidelines cause misinterpretations, class confusion, and boundary errors, systemic sources of inaccuracy rather than mere procedural flaws. While “Insufficient guidance” can also lead to inter-annotator variability (\textit{e.g.,} consistency errors), it is primarily categorised under accuracy because the root cause lies in inadequate or ambiguous annotation instructions that lead annotators to produce incorrect or imprecise labels relative to the intended ground truth. 

\textbf{(2.5) Bias-Driven Errors:}  This error refers to systematic deviations in annotation outcomes caused by human, contextual, or automation-induced biases that distort the accurate representation of real-world phenomena. Systematic deviations arise from cognitive or automation-induced bias. \textit{“Subsets systematically mislabeled (\textit{e.g.,}., red cars)… supervisors might have interpreted differently.”} (ID2) and \textit{“Automation tools rubber-stamped labels… annotators over-trusted tool outputs.”} (ID6) 
Bias-driven errors arose from unbalanced datasets, misaligned taxonomies, and overreliance on automated labelling, compounded by human cognitive biases, automation overtrust, and inconsistent quality assurance, which reinforced systemic data skew and unfairness (ID9, ID11, ID15). Overreliance on automated pre-labelling and unbalanced samples (ID8–ID15) propagates bias into perception models, harming fairness and trustworthiness.

\textbf{(3) Consistency Errors.} Under this theme, we cover both procedural and data oriented consistency errors. 
While accuracy concerns whether a label is correct with respect to a known ground truth, consistency concerns whether multiple annotators or tools would label the same instance in the same way.

\textbf{(3.1) Inter-Annotator Disagreement:} This error refers to the variation in labels assigned by different annotators to the same data instance, caused by subjective judgments, ambiguous guidelines, or inconsistent interpretation, resulting in annotation noise and reduced dataset reliability.  As one participant emphasised, \textit{“Yes, consensus is essential. If multiple annotators work on the same data, their outputs should align to ensure accuracy.”} (ID1) Another added, \textit{“Different annotators … they will subjectively annotate the object depending on distance or shape … then we need a second person to correct this.”} (ID18)  Human judgment variability, vague definitions of relevance or environmental context (ID3, ID6–ID7), and cost-driven limits on extensive review cycles (ID9, ID15) increase annotation noise and reduce dataset reliability, requiring additional reconciliation efforts before model training.

\textbf{(3.2) Ambiguous Instructions:} This error refers to unclear,  complex, or contradictory annotation guidelines that cause annotators to interpret labelling rules differently, leading to inconsistent annotations and reduced data quality.  One expert explained, \textit{“Clients often provide guidelines, but we frequently revise them to improve efficiency. I advocate for simplification annotators should not have to read 140 pages to understand the guidelines.”} (ID2) Similarly, another observed, \textit{“Determining which pedestrians are ‘relevant’ was unclear; guidelines did not define it strictly, leading to divergent [inconsistent] interpretations.”} (ID6)
Ambiguous instructions, such as overcomplicated or overly descriptive guidelines (ID1–ID3, ID5), missing contextual criteria (\textit{e.g.}, occlusions, lighting, traffic density) (ID7–ID9), lack of communication channels for clarifications (ID13), and unrealistic attempts to eliminate all ambiguity without iterative updates (ID16) amplify disagreement and rework, as annotators interpret rules differently under time pressure or limited supervision.

\textbf{(3.3) Lack of Purpose Knowledge:}  This error refers to annotators’ limited understanding of the dataset’s intended use or downstream model goals, leading to misaligned labelling decisions and inconsistent relevance judgments. Annotators often lack awareness of the dataset’s end-use, leading to divergent judgments and inconsistent relevance assessments. As one participant stated, \textit{“The annotators sometimes don’t understand why they are annotating certain things or how it ties into the overall purpose. They often say, ‘If we knew what this was supposed to train for, we could do it better.’”} (ID4) Another noted, \textit{“Requirements pass through multiple stakeholders, and annotators often don’t know the true purpose of labelling. This loss of context leads to inconsistent quality.”} (ID6)
 Poor communication of downstream objectives (ID8, ID11), insufficient domain expertise (ID12), and limited onboarding for new annotators result in inconsistent boundary choices, missing edge-case handling, and reduced overall alignment with model requirements.

\textbf{(3.4) Misaligned Hand-offs:} This error refers to inconsistencies and information loss that occur when annotation tasks, requirements, or guidelines are transferred between teams or stakeholders without proper documentation or communication, resulting in fragmented, inconsistent datasets.
Inconsistencies often emerge during transitions between client, QA, and annotation teams. \textit{“OEMs define requirements, but these may not be directly shared with suppliers. By the time they reach annotators, the instructions are incomplete or altered, creating inconsistencies.”} (ID6) Another participant added, \textit{“If a team does annotation internally instead of outsourcing, they might not document it properly. Instead, they’ll convey details in meetings, and that information can be very temporary, easily forgotten.”} (ID11)
Under-documented communication, missing shared tracking tools, and inconsistent rule interpretation across multiple vendors (ID4–ID6, ID8, ID11, ID14)  lead to fragmented datasets and require post-hoc harmonisation to ensure uniformity before model deployment.

\textbf{(3.5) Limited Review \& Logging:} This error refers to the absence of consistent quality checks and detailed record keeping during the annotation process, which prevents error detection, traceability, and reproducibility of annotation decisions. Scarce QA resources and weak process traceability leave inconsistencies undetected. One participant explained, \textit{“Given the vast data, manual review is not economically viable. With too few QA resources, errors remain undetected.”} (ID6) Another added, \textit{“Annotation companies rely on manual reviewing … they said they want 95\% accuracy, but it was all done manually … automated QA was lacking.”} (ID18) Insufficient automated validation (ID10, ID13), limited sampling or inter-annotator agreement checks (ID14–ID17), and high costs of independent verification (ID15–ID16)  result, corrupted or inconsistent labels (\textit{e.g.,}., bounding-box drift or missing objects) . These errors propagate into training data, lowering model stability and generalisation.

\textbf{(3.6) Lack of Frameworks and Standards:} This error refers to the absence of unified annotation protocols, ontologies, or benchmarking criteria across projects or organisations, leading to inconsistent practices, poor comparability, and reduced reproducibility of annotated datasets. The absence of shared annotation standards or tool benchmarks leads to inconsistent practices across organisations. One participant summarised, \textit{“One of the biggest challenges is the lack of standardisation in the annotation market… It is difficult to compare different annotation providers because there is no universal benchmark.”} (ID3) Similarly, another noted, \textit{“There is no common standard … every company follows their own guideline … sometimes datasets lack information about how annotations were done.”} (ID18)
Missing international standards (ID6, ID9, ID13), inconsistent adherence to ISO/UNECE guidelines (ID10, ID15), and incomplete documentation in datasets limit comparability, benchmarking, and reproducibility across AIePS.

\textbf{(3.7) Cross-Modality Misalignment:}   This error refers to spatial or temporal inconsistencies between annotations across different sensor modalities (\textit{e.g.,} camera, LiDAR, radar), caused by calibration errors or synchronisation gaps, leading to inaccurate data fusion and degraded perception performance. Misalignment between sensor modalities (\textit{e.g.,}., camera–LiDAR–radar) results in inconsistent spatial annotations. As one participant highlighted, \textit{“When multiple sensors record the same scene, they need to be registered together… You need calibration to make sure they align.”} (ID10) Another added, \textit{“We projected annotations from point cloud onto the image plane and checked alignment… sometimes the projection didn’t capture the object correctly.”} (ID18)
 Temporal sampling differences (ID13, ID17), inadequate calibration protocols (ID9–ID10), and missing 3D-to-2D projection validation distort object positioning and tracking accuracy across modalities, impairing fusion-based perception and sensor reliability.

 \begin{table*}[!h]
\centering
\scriptsize
\setlength{\tabcolsep}{10pt}
\renewcommand{\arraystretch}{1.2}
\caption{Expert validation outcomes, industrial utility, and implications for the taxonomy of data annotation errors}
\label{tab:RQ2_validation}

\begin{tabularx}{\textwidth}{
    >{\RaggedRight\arraybackslash}p{2.4cm}  
    Y                                        
    Y                                        
    >{\RaggedRight\arraybackslash}p{2.8cm}   
    Y                                        
}
\toprule
\rowcolor{thead}   
\thead{Theme} &
\thead{Validation Outcomes} &
\raisebox{1.2ex}{\thead{Representative \\ Comments \\ (OEM / Tier-1 / Tier-2)}} &
\thead{Industrial Utility} &
\thead{Implications} \\
\midrule

\textbf{Usefulness} &
High relevance for structuring and diagnosing annotation failures; recognised as a cross-industry quality reference. &
\qcell{OEM: Can be used as a failure-mode catalogue.}\par
\qcell{Tier-1: Helpful when setting up new annotation pipelines.}\par
\qcell{Tier-2: Useful as a checklist and for systematic analytics.} &
\textbf{Strengthens QA traceability.} Enables structured failure diagnosis and cross-team quality review. &
Validated taxonomy’s generalisability; retained as a framework for audits and supplier quality reviews. \\

\textbf{Practical Experience} &
The taxonomy is clear and comprehensive, suggesting the merging of overlapping categories. &
\qcell{Tier-2: Human vs automation not enough—add tool and guideline causes.} &
\textbf{Improves clarity of defect reporting.} Supports consistent error documentation across annotation teams. &
Expanded causes to include “tool-related” and “guideline-related” error types. \\

\textbf{Relevance} &
Consensus on three dimensions: completeness, accuracy, and consistency, aligned with ISO-based quality standards. &
\qcell{OEM: These are the same dimensions we face daily.} &
\textbf{Ensures interpretability in audits.} Aligns with existing industry QA frameworks. &
Maintained three main categories; clarified overlaps through refined definitions. \\

\textbf{Suggestions} &
Encouraged merging sub-items, simplifying structure, and adding quantitative risk indicators. &
\qcell{Tier-2: Add a risk-assessment matrix linking severity and probability.} &
\textbf{Enables risk-based prioritisation.} Supports safety-critical assessment of annotation errors. &
Planned extension of taxonomy with severity–frequency mapping (akin to FMEA scoring). \\

\textbf{Use Case Scenarios} &
Realistic applications include training, QA reviews, and root-cause analysis. &
\qcell{OEM: Use for checking perception-system root causes.}\par
\qcell{Tier-1: Cheat sheet for reviewing annotation processes.} &
\textbf{Enhances training and collaboration.} Standardises terminology across suppliers and OEMs. &
Evidence of immediate applicability; supports integration into QA tools and training programs. \\

\textbf{Errors and Integration} &
Systematic mislabelling, vague instructions, and missing edge cases are considered most damaging. &
\qcell{Tier-2: Systematic errors lead to ML models learning incorrect patterns.} &
\textbf{Supports model reliability and safety.} Enables early identification of critical annotation faults. &
Prioritised these error types in the final taxonomy; marked as high-severity recurrent issues. \\

\bottomrule
\end{tabularx}
\end{table*}

\subsection{RQ2: Experts' Validation of the Data Annotation Errors Taxonomy}
\label{sec:validation}

\textbf{Validation Purpose and Scope.} 
The validation focused on confirming the accuracy, relevance, and industrial applicability of the identified error categories rather than generating new data. Thats the reason the validation involved four experts ($1$ OEM, $1$ Tier-$1$, $2$ Tier-2); these participants were senior practitioners directly involved in data annotation governance and quality assurance, selected for their domain expertise and cross-organisational perspective. While all experts were based in European automotive contexts, their organisations operate globally, ensuring exposure to international standards and practices. This confirmatory and triangulated approach, grounded in design science and empirical software engineering principles~\cite{hevner2004design, runeson2009guidelines}, ensured the taxonomy’s practical relevance and completeness within real-world annotation workflows. The aim of this validation was depth and expert triangulation rather than statistical generalisation. 

\textbf{Validation Approach.}
Experts reviewed the data annotation error taxonomy using a structured questionnaire covering four dimensions: (1) their background and experience in annotation workflows, (2) the taxonomy’s usefulness and relevance, (3) its clarity, categorisation, and improvement areas, and (4) its practical applicability within supply chain processes. This systematic design ensured a balanced evaluation of both conceptual soundness and industrial usability (see \href{https://dataverse.harvard.edu/previewurl.xhtml?token=da9478c0-7653-48cb-bc5f-b8ae1d0810e8}{\textcolor{blue}{→ Structured Validation Questions}}). The  (Table \ref{tab:RQ2_validation})  shows details of the complete validation responses.

\textbf{(1) Expert Profiles: Starting Questions (Q1–Q4).} The validation involved experts, including a System Safety Expert (OEM), Machine Learning Engineer (Tier-$1$), Director of Customer Services (Tier-$2$), and Perception Expert (Tier-$2$). Each possessed $4–5$ years of experience in annotation workflows and AIePS development, with direct roles in data collection, annotation, and AI system integration. 

\textbf{(2) Usefulness, Relevance, and Applicability (Q5–Q16).}
Across all experts, the taxonomy was perceived as highly relevant, practical, and directly applicable within industrial settings for structuring, diagnosing, and improving data annotation workflows.

- The \textbf{OEM system safety expert} viewed the taxonomy as a \textit{“failure-mode catalogue”} analogous to FMEA (Failure Mode and Effects Analysis), helping identify and trace data-related weaknesses in perception-system development. It was considered particularly valuable for root-cause analysis, supplier quality reviews, and training of perception validation teams. The expert also noted that it supports the diagnosis of perception failures (\textit{e.g.,} inconsistent object recognition or missed detections) and facilitates cross-team communication among the safety, perception, and data quality units.

- The \textbf{Tier-$1$ ML engineer} described the taxonomy as a practical \textit{“checklist”} for setting up new annotation pipelines, reviewing guidelines, and validating existing annotation processes. It was also seen as a “cheat sheet” for systematically verifying completeness, identifying recurring issues, and ensuring that all major error sources are considered during the design or revision of data annotation workflows.

 - The \textbf{Tier-$2$ director of customer services  and perception expert} emphasised that the taxonomy supports \textit{“systematic analytics of annotation errors”} and improves \textit{“onboarding for new projects.”} They found it particularly useful for structuring internal quality assurance (QA), facilitating client discussions, and aligning understanding between annotation providers and customers. Perception expert further highlighted its potential as onboarding material for new employees and clients, and as a reference during guideline optimization and annotation tool configuration.

\textbf{(3) Reflection on  Clarity, and Improvements (Q9–Q12).}
All experts agreed that the three overarching quality dimensions, completeness, accuracy, and consistency, effectively capture the main data quality challenges observed in industrial contexts. The OEM noted that these dimensions align with established quality and safety standards such as ISO $21448$, while the Tier-$1$ expert emphasised simplifying and consolidating categories to facilitate adoption, a recommendation incorporated into the final taxonomy refinement. Tier-$2$ experts proposed distinguishing more clearly between human-, automation, and tool-induced errors, which informed the cause impact grouping in our analysis. Collectively, experts encouraged extending the taxonomy through studies of real-world failure cases and linking each error type with measurable severity indicators such as frequency or safety impact.


\textbf{(4) Preliminary Risk Categorization of Errors}. The experts consistently emphasised high-risk errors such as systematic mislabeling, incomplete or edge-case omissions, inconsistent annotations, and schema mismatches, all of which align with the taxonomy’s 18 core errors. In contrast, several low-frequency but practically significant errors also emerged, including cross-modality misalignment, low-quality pre-annotations, limited feedback loops, and human fatigue. While less frequently mentioned, these emergent cases reveal evolving challenges arising from automation, organisational constraints, and human variability, indicating where future quality assurance efforts and process improvements should focus. (see: \href{https://dataverse.harvard.edu/previewurl.xhtml?token=da9478c0-7653-48cb-bc5f-b8ae1d0810e8}{\textcolor{blue}{→ Table 3- Expert Classification of Annotation Errors by Risk Level.}})



\section{Threats to Validity}

Following Runeson \textit{et al.}~\cite{runeson2009guidelines} and Maxwell~\cite{maxwell_validity}, potential validity threats and mitigation measures are outlined below.

\textbf{Construct Validity.}
To ensure participant relevance, experts were purposefully selected from OEMs, Tier-$1$, Tier-$2$, and research institutes. The interview protocol underwent internal review and two pilot tests for clarity and completeness. Key terms (\textit{e.g.,} data annotation, perception systems) were clearly defined and supported with visuals to avoid misinterpretation. For expert validation, participants received structured briefings and could seek clarification before evaluation. Researcher bias was minimised through team-based reflexive discussions and peer debriefing.

\textbf{Internal Validity.}
To mitigate personal or organisational bias, participants were drawn from six automotive companies and four research institutes across multiple supply chain tiers. Interviews were recorded, transcribed, and participant-verified for accuracy. Dual independent coding yielded $\kappa = 0.8$ (Cohen’s Kappa), with discrepancies resolved during weekly meetings from May to September 2025. Three collaborative workshops with researchers and industry partners refined the themes, while triangulation between interview (RQ1) and validation (RQ2) results enhanced conceptual and practical credibility.

\textbf{External Validity.}
Although not statistically generalisable, the study achieves analytical generalisation through diverse participants, $20$ experts from $11$ organisations across two European countries and the UK. Coverage across OEMs, Tier-$1$/Tier-$2$ suppliers, and research institutions, combined with alignment to international standards (ISO/IEC~$5259$, IEEE~$P2801$, ISO~$26262$, SAE~J$3016$), strengthens transferability and ensures lifecycle representativeness within AIePS.

\textbf{Reliability.}
To ensure transparency and replicability, all instruments (interview guide, codebook, validation questionnaire) are openly available. Interviews were conducted with informed consent, anonymised in accordance with GDPR, and securely stored. All research steps, from data collection to coding logic, were systematically documented to support replication and independent verification.

\section{Conclusions and Discussion}
This study presents a structured taxonomy of $18$ data annotation errors in AI-enabled automotive perception, integrating theoretical data-quality concepts with industrial insights. Grounded in the dimensions of completeness, accuracy, and consistency.

\begin{figure}[htbp]  
    \centering
    \includegraphics[width=0.5\textwidth]{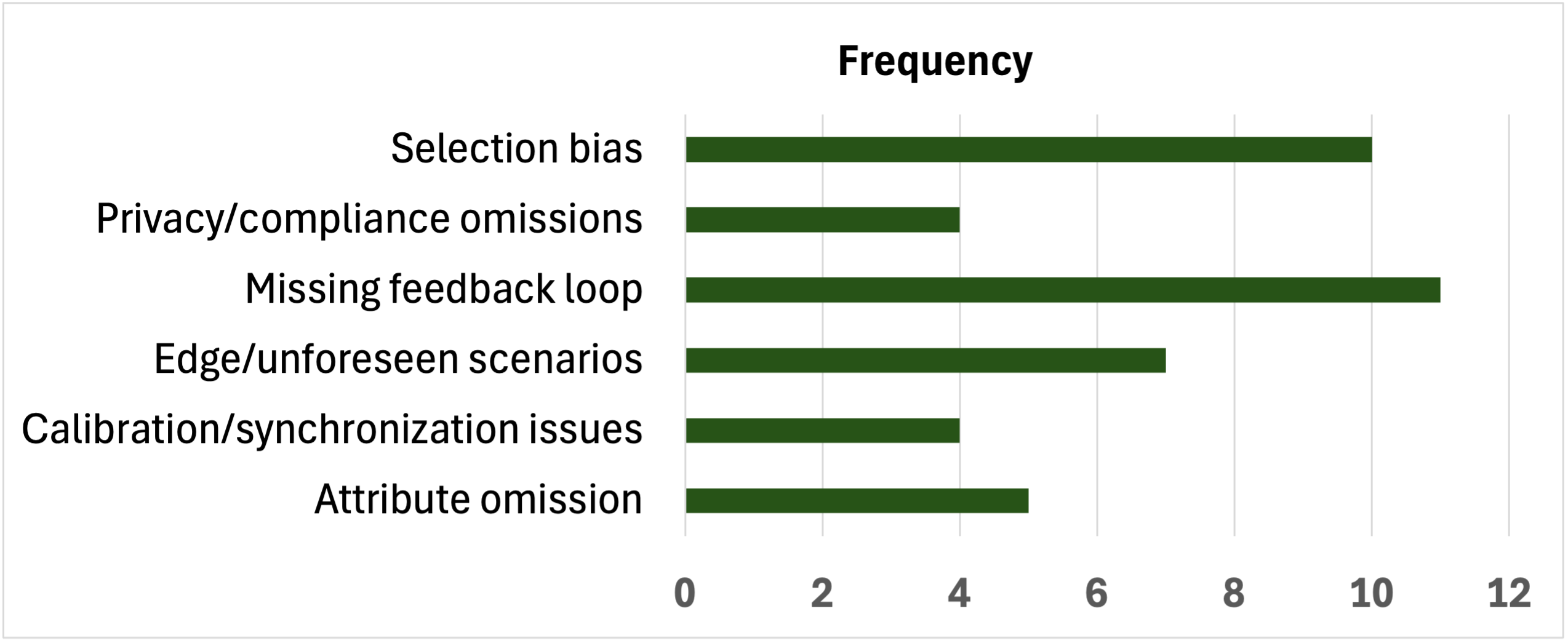} 
    \caption{ High mentioned completeness errors. }
   \label{fig:method1}
\end{figure}

\textbf{Data Annotation Errors and the State of the Art.} Completeness errors capture the underrepresentation of critical driving scenarios and incomplete information capture in perception datasets~\cite{wang2024survey,bachute2021autonomous}. Traditional issues such as \textit{edge-case omissions}, \textit{selection bias}, and \textit{attribute omission} align with known “long-tail” challenges~\cite{heyn2023automotive,beck2023quality,klie2024analyzing}. Yet, our taxonomy extends this view by introducing novel dimensions such as \textit{privacy/compliance omissions}, \textit{missing feedback loops}, and \textit{sensor synchronisation issues}~\cite{wac2023capturing,ruggeri2024let,zhong2022detecting,chen2022road}. These additions highlight that completeness is not merely a matter of dataset size but of regulatory, temporal, and multimodal adequacy across sensors and stakeholders. The Figure~\ref{fig:method1} highlights the most frequently mentioned completeness-related errors in the study. Missing feedback loop and selection bias were the most common issues, followed by edge/unforeseen scenarios. Less frequent but notable errors included attribute omission, privacy/compliance omissions, and calibration/synchronization issues. Overall, the findings suggest that systemic gaps, particularly in feedback and data selection, are the main contributors to completeness errors.

\begin{figure}[htbp]  
    \centering
    \includegraphics[width=0.5\textwidth]{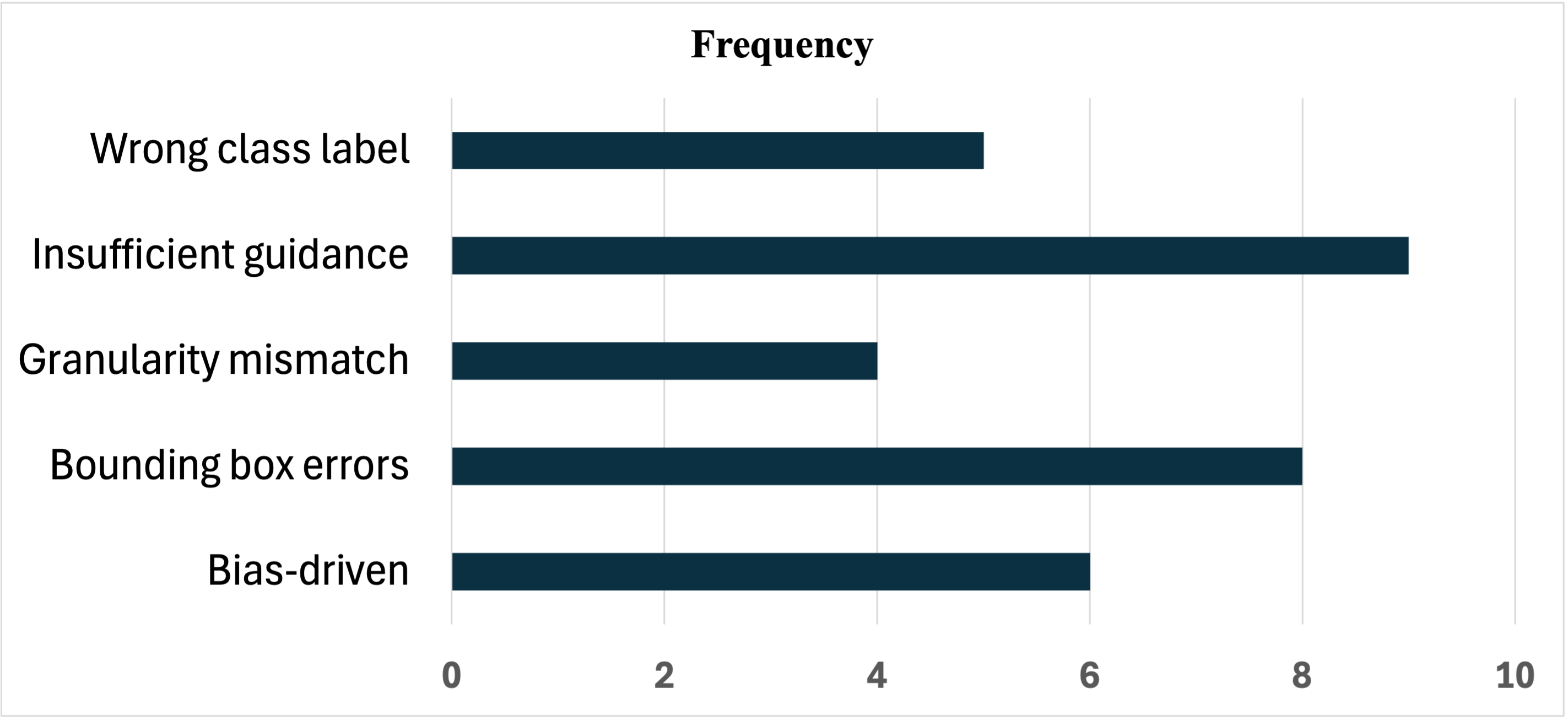} 
    \caption{ High mentioned accuracy errors. }
   \label{fig:method2}
\end{figure}

Accuracy errors represent distortions between the intended ground truth and the final annotation. Common cases such as \textit{wrong class labels}, \textit{bounding-box errors}, and \textit{granularity mismatches} remain frequent causes of label noise~\cite{penquitt2025label,schubert2024identifying,klugmann2024no}. At the same time, industrial evidence revealed human automation interaction issues \textit{automation over trust}, \textit{insufficient guidance}, and \textit{lack of calibration feedback} that amplify inaccuracies through over-reliance on automated pre-labelling tools~\cite{beck2023quality,ruggeri2024let,watson2023augmented}. These findings bridge technical and socio-organisational sources of label errors, showing that accuracy degradation often stems from both data limitations and human decision making biases.
Figure~\ref{fig:method2} presents the most frequently mentioned accuracy related errors in the study. Insufficient guidance was the most commonly reported issue, followed by bounding box errors and wrong class labels, each occurring frequently. Bias driven errors were also notable, while granularity mismatch appeared less often. Overall, the findings suggest that accuracy problems stem largely from unclear annotation instructions and technical labelling inconsistencies, underscoring the need for clearer guidelines and improved annotation precision.

\begin{figure}[htbp]  
    \centering
    \includegraphics[width=0.5\textwidth]{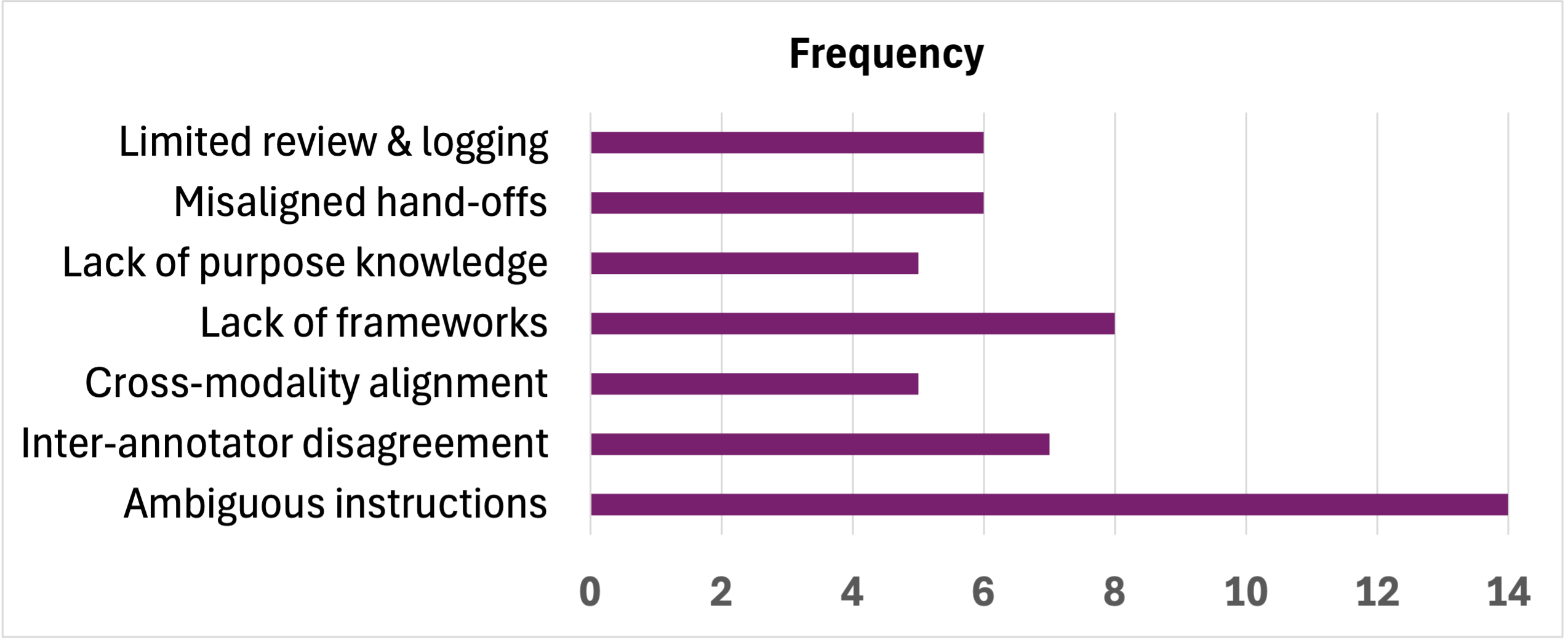} 
    \caption{ High mentioned consistency errors. }
   \label{fig:consistency}
\end{figure}

Consistency errors denote deviations in labelling uniformity across annotators, projects, and modalities~\cite{klie2024analyzing,beck2023quality,van2024deciphering}. While inconsistent instructions and inadequate calibration remain established concerns, our taxonomy identifies process-level consistency failures unique to distributed industrial pipelines such as \textit{misaligned hand-offs}, \textit{cross-modality misalignment}, and lack of \textit{standardisation frameworks}~\cite{zhong2022detecting,borg2023ergo,wac2023capturing}. Many of these issues stem from socio-technical misalignments, including ambiguous guidelines, workload variability, and lack of motivational factors and annotator fatigue~\cite{ohman2020challenges,beck2023quality}. Addressing such issues thus requires not only better annotation tools but also systematic processes such as feedback loops, clearer organisational responsibilities, and continuous cross-team calibration. As shown in the Figure~\ref{fig:consistency}, ambiguous instructions were the most frequently reported consistency-related error in our study, followed by inter-annotator disagreement and lack of frameworks or standards. Other notable issues, such as misaligned hand-offs and limited review and logging, point to process level errors, while lack of purpose knowledge and cross-modality alignment appeared less often but remain relevant. We observed that annotation inconsistency arises from both human variability and systemic gaps in process standardisation and guidance. 

\textbf{Data Annotation Errors and Quality Degradation.}
The study demonstrates that annotation errors are not random or isolated incidents but systemic drivers of downstream quality degradation in perception models. Completeness errors, such as missing feedback loops or edge case omissions, reduce dataset representativeness and weaken model generalisation~\cite{wang2024survey,bachute2021autonomous}. Accuracy-related errors, including mislabeling, bounding-box misplacement, and granularity mismatches, distort learning signals and reduce predictive reliability~\cite{beck2023quality,klie2024analyzing,penquitt2025label}. Likewise, consistency issues from inter-annotator disagreement to cross-modality misalignment introduce latent noise that propagates through training and deployment~\cite{heyn2023automotive,ohman2020challenges}. Together, these defects can cascade throughout the AI lifecycle, undermining performance in safety-critical metrics such as false positives, missed detections, and perception uncertainty. Grounded in the established principles of \textit{completeness}, \textit{accuracy}, and \textit{consistency}~\cite{wand1996anchoring,pipino2002data,batini2009methodologies}, the taxonomy operationalises abstract data-quality constructs into concrete, auditable categories. This provides a data-centric assurance foundation aligned with the broader trustworthy-AI agenda, emphasising traceability, explainability, and governance in line with EU and ISO standards.

\textbf{Emergent Low-Frequency Annotation Errors.}
Beyond the 18 high-frequency error types, our analysis also revealed several low-frequency but practically significant error types across the completeness, accuracy, and consistency dimensions. From a completeness perspective, these low-mention cases point to partial or uneven data coverage, including missing modalities, skipped frames, and restrictive dataset curation driven by technical, legal, or organisational constraints~\cite{wang2024survey,zhong2022detecting,wac2023capturing,ruggeri2024let}. We also found automation related omissions, for instance, when foundation models overlook rare or privacy sensitive instances and quality lapses such as incomplete pre-annotations or malfunction induced data loss~\cite{watson2023augmented,beck2023quality}.
For accuracy, the low-frequency issues reflect small spatial or temporal misalignments, hybrid or automated labelling errors, and human factors such as fatigue, bias, and overreliance on automation~\cite{penquitt2025label,schubert2024identifying,shen2020automated,klugmann2024no,ohman2020challenges,wang2024survey}. Even minor inaccuracies in annotation precision can propagate through training pipelines, degrading perception reliability.
Under consistency, low-frequency cases reveal gradual or structural drifts in annotation uniformity across projects or over time~\cite{klie2024analyzing,beck2023quality}. These arise from evolving guidelines, automation carryover errors, and mismatches between annotators’ understanding and supervisors’ expectations~\cite{watson2023augmented,shen2020automated}. Variations in workload, motivation, and team calibration, along with limited onboarding or expert oversight, further amplify such inconsistencies~\cite{darji2024challenges,ohman2020challenges}.
Collectively, these low-frequency but conceptually important errors reveal emerging challenges in modern annotation pipelines, where automation, organisational constraints, and human variability intersect to shape data quality beyond the dominant error categories captured in the taxonomy.

\textbf{Industrial Validation and Implications.}
Industry experts across OEMs, Tier-$1$, and Tier-$2$ suppliers validated the taxonomy’s clarity, relevance, and operational value. They characterised it as a “failure-mode catalogue” analogous to FMEA, an interpretation absent in prior literature. The taxonomy enables direct traceability from perception anomalies (\textit{e.g.,}., false positives, missed detections) to upstream data defects (\textit{e.g.,}., incomplete coverage, misaligned hand-offs). It provides a shared vocabulary for diagnosing whether an issue originates from human annotation, automation, or tooling, thereby improving cross-tier communication and quality control. Experts also recognised its educational value for onboarding, internal QA reviews, and continuous improvement, and suggested integrating it into annotation dashboards to institutionalise a data-quality culture. 

\textbf{Trustworthy AI Developemnt and Data Governance Implications.
}
The data annotation errors taxonomy contributes to the broader trustworthy AI agenda by linking data quality, process transparency, and accountability across the AI systems developemnt lifecycle. It supports traceability of errors from perception anomalies to underlying annotation causes, an essential capability for compliance with emerging standards such as ISO/IEC~$5259$, IEEE~$P2801$, ISO~$26262$, $SAE$ ~$J3016$ lifecycle assurance with these frameworks. The taxonomy provides a practical foundation for governance-by-design, where annotation quality becomes an auditable and traceable element of AI system certification. In doing so, it operationalises fairness, transparency, and accountability as measurable engineering properties rather than abstract ethical goals.

\textbf{Implications for the AI-Enabled System Development Lifecycle.}
The taxonomy influences all phases of AI-enabled system development by embedding data quality considerations into engineering workflows. During \emph{data collection and curation}, it improves dataset representativeness by identifying completeness-related gaps. In \emph{annotation and tooling}, it guides workflow design to mitigate human–automation drift and reinforce auditability. For \emph{model training and evaluation}, it provides traceability between annotation noise and model reliability, informing retraining priorities. Within \emph{verification and validation (V\&V)}, it supports safety traceability by linking perception errors to underlying data issues. In \emph{requirements engineering}, it formalises data-specific criteria such as inter-annotator agreement thresholds, while in \emph{quality assurance and maintenance}, it establishes a foundation for audits, supplier evaluations, and continuous improvement. These cross-cutting implications position the taxonomy as a unifying framework aligning data governance, model assurance, and system level quality management across the AI lifecycle.

\textbf{Implications for Research and Practice.} Beyond AIePS developemnt life cycle integration, the taxonomy also offers broader implications for both research and industrial practice. For research purposes, this study establishes an empirically grounded taxonomy that systematises how data annotation errors manifest in the development of AIePS. It highlights the need for future research to move beyond isolated notions of “data quality” toward integrated frameworks linking annotation processes, organisational factors, and model performance. The taxonomy can serve as a foundation for quantitative studies measuring the impact of specific error types on model outcomes, as well as for developing automated tools that detect or predict annotation errors. For practice, the taxonomy provides practitioners with a structured lens for diagnosing and preventing quality issues throughout the annotation lifecycle. It can inform the design of quality assurance checklists, onboarding materials, and supplier evaluation criteria, fostering shared terminology across OEMs, Tier-1s, and annotation vendors. By identifying high-impact error types such as missing feedback loops, insufficient guidance, and weak quality control organisations can prioritise interventions that strengthen both annotation quality and governance. Moreover, the taxonomy encourages proactive error monitoring and the integration of feedback into existing annotation workflows, enabling continuous improvement. It also supports traceability by linking annotation errors to downstream perception failures, enhancing transparency in safety-critical development. Finally, it establishes a conceptual foundation for data governance standards and policy frameworks, ensuring that annotation quality is treated as a first-class element of system assurance rather than a peripheral task.

\textbf{Future Work.}
 Future research should expand validation across diverse geographical and industrial contexts to strengthen generalisability. Another critical direction is the development of semi-automated quality assurance systems that apply the taxonomy for real-time monitoring, drift detection, and prioritised correction, thus enabling continuous data assurance throughout the AI lifecycle.
Furthermore, quantitative studies correlating the frequency and severity of specific error types with model performance metrics would deepen understanding of their practical impact. Integrating severity probability matrices, similar to FMEA (Failure Mode and Effects Analysis) scoring, can support risk-based prioritisation in industrial annotation workflows.
By embedding such automated mechanisms into annotation tools and MLOps pipelines, future work can transform the taxonomy from a diagnostic artefact into a proactive engineering instrument for lifecycle assurance in SE4AI.

\textbf{ACKNOWLEDGMENTS} This research is supported by Vinnova, Program Fordons strategisk Forskning och Innovation (FFI), Project FAMER (2023-00771).

\bibliographystyle{ACM-Reference-Format}
\bibliography{refferences}


\begin{thebibliography}{41}


\ifx \showCODEN    \undefined \def \showCODEN     #1{\unskip}     \fi
\ifx \showISBNx    \undefined \def \showISBNx     #1{\unskip}     \fi
\ifx \showISBNxiii \undefined \def \showISBNxiii  #1{\unskip}     \fi
\ifx \showISSN     \undefined \def \showISSN      #1{\unskip}     \fi
\ifx \showLCCN     \undefined \def \showLCCN      #1{\unskip}     \fi
\ifx \shownote     \undefined \def \shownote      #1{#1}          \fi
\ifx \showarticletitle \undefined \def \showarticletitle #1{#1}   \fi
\ifx \showURL      \undefined \def \showURL       {\relax}        \fi
\providecommand\bibfield[2]{#2}
\providecommand\bibinfo[2]{#2}
\providecommand\natexlab[1]{#1}
\providecommand\showeprint[2][]{arXiv:#2}

\bibitem[Bachute and Subhedar(2021)]%
        {bachute2021autonomous}
\bibfield{author}{\bibinfo{person}{Mrinal~R Bachute} {and} \bibinfo{person}{Javed~M Subhedar}.} \bibinfo{year}{2021}\natexlab{}.
\newblock \showarticletitle{Autonomous driving architectures: insights of machine learning and deep learning algorithms}.
\newblock \bibinfo{journal}{\emph{Machine Learning with Applications}}  \bibinfo{volume}{6} (\bibinfo{year}{2021}), \bibinfo{pages}{100164}.
\newblock


\bibitem[Batini et~al\mbox{.}(2009)]%
        {batini2009methodologies}
\bibfield{author}{\bibinfo{person}{Carlo Batini}, \bibinfo{person}{Cinzia Cappiello}, \bibinfo{person}{Chiara Francalanci}, {and} \bibinfo{person}{Andrea Maurino}.} \bibinfo{year}{2009}\natexlab{}.
\newblock \showarticletitle{Methodologies for data quality assessment and improvement}.
\newblock \bibinfo{journal}{\emph{ACM Computing Surveys (CSUR)}} \bibinfo{volume}{41}, \bibinfo{number}{3} (\bibinfo{year}{2009}), \bibinfo{pages}{1--52}.
\newblock
\href{https://doi.org/10.1145/1541880.1541883}{doi:\nolinkurl{10.1145/1541880.1541883}}


\bibitem[Beck(2023)]%
        {beck2023quality}
\bibfield{author}{\bibinfo{person}{Jacob Beck}.} \bibinfo{year}{2023}\natexlab{}.
\newblock \showarticletitle{Quality aspects of annotated data: A research synthesis}.
\newblock \bibinfo{journal}{\emph{AStA Wirtschafts-und Sozialstatistisches Archiv}} \bibinfo{volume}{17}, \bibinfo{number}{3} (\bibinfo{year}{2023}), \bibinfo{pages}{331--353}.
\newblock


\bibitem[Borg et~al\mbox{.}(2023)]%
        {borg2023ergo}
\bibfield{author}{\bibinfo{person}{Markus Borg}, \bibinfo{person}{Jens Henriksson}, \bibinfo{person}{Kasper Socha}, \bibinfo{person}{Olof Lennartsson}, \bibinfo{person}{Elias Sonnsj{\"o}~L{\"o}negren}, \bibinfo{person}{Thanh Bui}, \bibinfo{person}{Piotr Tomaszewski}, \bibinfo{person}{Sankar~Raman Sathyamoorthy}, \bibinfo{person}{Sebastian Brink}, {and} \bibinfo{person}{Mahshid Helali~Moghadam}.} \bibinfo{year}{2023}\natexlab{}.
\newblock \showarticletitle{Ergo, SMIRK is safe: a safety case for a machine learning component in a pedestrian automatic emergency brake system}.
\newblock \bibinfo{journal}{\emph{Software quality journal}} \bibinfo{volume}{31}, \bibinfo{number}{2} (\bibinfo{year}{2023}), \bibinfo{pages}{335--403}.
\newblock


\bibitem[Braun and Clarke(2006)]%
        {braun2006using}
\bibfield{author}{\bibinfo{person}{Virginia Braun} {and} \bibinfo{person}{Victoria Clarke}.} \bibinfo{year}{2006}\natexlab{}.
\newblock \showarticletitle{Using thematic analysis in qualitative research}.
\newblock \bibinfo{journal}{\emph{Qualitative Research in Psychology}} \bibinfo{volume}{3}, \bibinfo{number}{2} (\bibinfo{year}{2006}), \bibinfo{pages}{77--101}.
\newblock


\bibitem[Chacon-Murguia and Prieto-Resendiz(2015)]%
        {chacon2015detecting}
\bibfield{author}{\bibinfo{person}{Mario~I Chacon-Murguia} {and} \bibinfo{person}{Claudia Prieto-Resendiz}.} \bibinfo{year}{2015}\natexlab{}.
\newblock \showarticletitle{Detecting driver drowsiness: A survey of system designs and technology}.
\newblock \bibinfo{journal}{\emph{IEEE Consumer Electronics Magazine}} \bibinfo{volume}{4}, \bibinfo{number}{4} (\bibinfo{year}{2015}), \bibinfo{pages}{107--119}.
\newblock


\bibitem[Chen et~al\mbox{.}(2022)]%
        {chen2022road}
\bibfield{author}{\bibinfo{person}{Rung-Ching Chen}, \bibinfo{person}{Vani~Suthamathi Saravanarajan}, \bibinfo{person}{Long-Sheng Chen}, {and} \bibinfo{person}{Hui Yu}.} \bibinfo{year}{2022}\natexlab{}.
\newblock \showarticletitle{Road segmentation and environment labeling for autonomous vehicles}.
\newblock \bibinfo{journal}{\emph{Applied Sciences}} \bibinfo{volume}{12}, \bibinfo{number}{14} (\bibinfo{year}{2022}), \bibinfo{pages}{7191}.
\newblock


\bibitem[Darji et~al\mbox{.}(2024)]%
        {darji2024challenges}
\bibfield{author}{\bibinfo{person}{Harshil Darji}, \bibinfo{person}{Jelena Mitrovi{\'c}}, {and} \bibinfo{person}{Michael Granitzer}.} \bibinfo{year}{2024}\natexlab{}.
\newblock \showarticletitle{Challenges and Considerations in Annotating Legal Data: A Comprehensive Overview}.
\newblock \bibinfo{journal}{\emph{arXiv preprint arXiv:2407.17503}} (\bibinfo{year}{2024}).
\newblock


\bibitem[Demrozi et~al\mbox{.}(2021)]%
        {demrozi2021towards}
\bibfield{author}{\bibinfo{person}{Florenc Demrozi}, \bibinfo{person}{Marin Jereghi}, {and} \bibinfo{person}{Graziano Pravadelli}.} \bibinfo{year}{2021}\natexlab{}.
\newblock \showarticletitle{Towards the automatic data annotation for human activity recognition based on wearables and BLE beacons}. In \bibinfo{booktitle}{\emph{2021 IEEE International Symposium on Inertial Sensors and Systems (INERTIAL)}}. IEEE, \bibinfo{pages}{1--4}.
\newblock


\bibitem[Dey and Lee(2023)]%
        {dey2023multi}
\bibfield{author}{\bibinfo{person}{Sangeeta Dey} {and} \bibinfo{person}{Seok-Won Lee}.} \bibinfo{year}{2023}\natexlab{}.
\newblock \showarticletitle{A Multi-layered collaborative framework for evidence-driven data requirements engineering for machine learning-based safety-critical systems}. In \bibinfo{booktitle}{\emph{Proceedings of the 38th ACM/SIGAPP Symposium on Applied Computing}}. \bibinfo{pages}{1404--1413}.
\newblock


\bibitem[Fredriksson et~al\mbox{.}(2020)]%
        {fredriksson2020data}
\bibfield{author}{\bibinfo{person}{Teodor Fredriksson}, \bibinfo{person}{David~Issa Mattos}, \bibinfo{person}{Jan Bosch}, {and} \bibinfo{person}{Helena~Holmstr{\"o}m Olsson}.} \bibinfo{year}{2020}\natexlab{}.
\newblock \showarticletitle{Data labeling: An empirical investigation into industrial challenges and mitigation strategies}. In \bibinfo{booktitle}{\emph{International Conference on Product-Focused Software Process Improvement}}. Springer, \bibinfo{pages}{202--216}.
\newblock


\bibitem[Galv{\~a}o and Huda(2023)]%
        {galvao2023pedestrian}
\bibfield{author}{\bibinfo{person}{Luiz~G Galv{\~a}o} {and} \bibinfo{person}{M~Nazmul Huda}.} \bibinfo{year}{2023}\natexlab{}.
\newblock \showarticletitle{Pedestrian and vehicle behaviour prediction in autonomous vehicle system—A review}.
\newblock \bibinfo{journal}{\emph{Expert Systems with Applications}} (\bibinfo{year}{2023}), \bibinfo{pages}{121983}.
\newblock


\bibitem[Habibullah et~al\mbox{.}(2023)]%
        {habibullah2023requirements}
\bibfield{author}{\bibinfo{person}{Khan~Mohammad Habibullah}, \bibinfo{person}{Hans-Martin Heyn}, \bibinfo{person}{Gregory Gay}, \bibinfo{person}{Jennifer Horkoff}, \bibinfo{person}{Eric Knauss}, \bibinfo{person}{Markus Borg}, \bibinfo{person}{Alessia Knauss}, \bibinfo{person}{H{\aa}kan Sivencrona}, {and} \bibinfo{person}{Jing Li}.} \bibinfo{year}{2023}\natexlab{}.
\newblock \showarticletitle{Requirements engineering for automotive perception systems: An interview study}. In \bibinfo{booktitle}{\emph{International Working Conference on Requirements Engineering: Foundation for Software Quality}}. Springer, \bibinfo{pages}{189--205}.
\newblock


\bibitem[Hevner et~al\mbox{.}(2004)]%
        {hevner2004design}
\bibfield{author}{\bibinfo{person}{Alan~R. Hevner}, \bibinfo{person}{Salvatore~T. March}, \bibinfo{person}{Jinsoo Park}, {and} \bibinfo{person}{Sudha Ram}.} \bibinfo{year}{2004}\natexlab{}.
\newblock \showarticletitle{Design Science in Information Systems Research}.
\newblock \bibinfo{journal}{\emph{MIS Quarterly}} \bibinfo{volume}{28}, \bibinfo{number}{1} (\bibinfo{year}{2004}), \bibinfo{pages}{75--105}.
\newblock
\href{https://doi.org/10.2307/25148625}{doi:\nolinkurl{10.2307/25148625}}


\bibitem[Heyn et~al\mbox{.}(2023)]%
        {heyn2023automotive}
\bibfield{author}{\bibinfo{person}{Hans-Martin Heyn}, \bibinfo{person}{Khan~Mohammad Habibullah}, \bibinfo{person}{Eric Knauss}, \bibinfo{person}{Jennifer Horkoff}, \bibinfo{person}{Markus Borg}, \bibinfo{person}{Alessia Knauss}, {and} \bibinfo{person}{Polly~Jing Li}.} \bibinfo{year}{2023}\natexlab{}.
\newblock \showarticletitle{Automotive perception software development: An empirical investigation into data, annotation, and ecosystem challenges}. In \bibinfo{booktitle}{\emph{2023 IEEE/ACM 2nd International Conference on AI Engineering--Software Engineering for AI (CAIN)}}. IEEE, \bibinfo{pages}{13--24}.
\newblock


\bibitem[Khattak et~al\mbox{.}(2021)]%
        {khattak2021taxonomy}
\bibfield{author}{\bibinfo{person}{Asad~J Khattak}, \bibinfo{person}{Numan Ahmad}, \bibinfo{person}{Behram Wali}, {and} \bibinfo{person}{Eric Dumbaugh}.} \bibinfo{year}{2021}\natexlab{}.
\newblock \showarticletitle{A taxonomy of driving errors and violations: Evidence from the naturalistic driving study}.
\newblock \bibinfo{journal}{\emph{Accident Analysis \& Prevention}}  \bibinfo{volume}{151} (\bibinfo{year}{2021}), \bibinfo{pages}{105873}.
\newblock


\bibitem[Klie et~al\mbox{.}(2024)]%
        {klie2024analyzing}
\bibfield{author}{\bibinfo{person}{Jan-Christoph Klie}, \bibinfo{person}{Richard~Eckart de Castilho}, {and} \bibinfo{person}{Iryna Gurevych}.} \bibinfo{year}{2024}\natexlab{}.
\newblock \showarticletitle{Analyzing dataset annotation quality management in the wild}.
\newblock \bibinfo{journal}{\emph{Computational Linguistics}} (\bibinfo{year}{2024}), \bibinfo{pages}{1--48}.
\newblock


\bibitem[Klugmann et~al\mbox{.}(2024)]%
        {klugmann2024no}
\bibfield{author}{\bibinfo{person}{Christopher Klugmann}, \bibinfo{person}{Rafid Mahmood}, \bibinfo{person}{Guruprasad Hegde}, \bibinfo{person}{Amit Kale}, {and} \bibinfo{person}{Daniel Kondermann}.} \bibinfo{year}{2024}\natexlab{}.
\newblock \showarticletitle{No Need to Sacrifice Data Quality for Quantity: Crowd-Informed Machine Annotation for Cost-Effective Understanding of Visual Data}.
\newblock \bibinfo{journal}{\emph{arXiv preprint arXiv:2409.00048}} (\bibinfo{year}{2024}).
\newblock


\bibitem[Kr{\"u}ger(2022)]%
        {kruger2022keynote}
\bibfield{author}{\bibinfo{person}{Frank Kr{\"u}ger}.} \bibinfo{year}{2022}\natexlab{}.
\newblock \showarticletitle{Keynote: Adventures in Annotation: Providing High Quality Labels for Supervised Machine Learning}. In \bibinfo{booktitle}{\emph{2022 IEEE International Conference on Pervasive Computing and Communications Workshops and other Affiliated Events (PerCom Workshops)}}. IEEE, \bibinfo{pages}{254--254}.
\newblock


\bibitem[Kukkala et~al\mbox{.}(2018)]%
        {kukkala2018advanced}
\bibfield{author}{\bibinfo{person}{Vipin~Kumar Kukkala}, \bibinfo{person}{Jordan Tunnell}, \bibinfo{person}{Sudeep Pasricha}, {and} \bibinfo{person}{Thomas Bradley}.} \bibinfo{year}{2018}\natexlab{}.
\newblock \showarticletitle{Advanced driver-assistance systems: A path toward autonomous vehicles}.
\newblock \bibinfo{journal}{\emph{IEEE Consumer Electronics Magazine}} \bibinfo{volume}{7}, \bibinfo{number}{5} (\bibinfo{year}{2018}), \bibinfo{pages}{18--25}.
\newblock


\bibitem[Maxwell(1992)]%
        {maxwell_validity}
\bibfield{author}{\bibinfo{person}{Joseph Maxwell}.} \bibinfo{year}{1992}\natexlab{}.
\newblock \showarticletitle{Understanding and validity in qualitative research}.
\newblock \bibinfo{journal}{\emph{Harvard Educational Review}} \bibinfo{volume}{62}, \bibinfo{number}{3} (\bibinfo{year}{1992}), \bibinfo{pages}{279--301}.
\newblock


\bibitem[Mohammedali and Adam(2023)]%
        {mohammedali2023influence}
\bibfield{author}{\bibinfo{person}{Maab Mohammedali} {and} \bibinfo{person}{Muntasir Adam}.} \bibinfo{year}{2023}\natexlab{}.
\newblock \showarticletitle{The influence of data annotation process requirements on performance criteria of ML models}.
\newblock \bibinfo{journal}{\emph{Gothenburg University Library}} (\bibinfo{year}{2023}).
\newblock


\bibitem[Najafi and Boukerche(2024)]%
        {najafi2024performance}
\bibfield{author}{\bibinfo{person}{Alireza Najafi} {and} \bibinfo{person}{Azzedine Boukerche}.} \bibinfo{year}{2024}\natexlab{}.
\newblock \showarticletitle{On The Performance of Perception Systems of Autonomous Vehicles}. In \bibinfo{booktitle}{\emph{ICC 2024-IEEE International Conference on Communications}}. IEEE, \bibinfo{pages}{5365--5370}.
\newblock


\bibitem[{\"O}hman(2020)]%
        {ohman2020challenges}
\bibfield{author}{\bibinfo{person}{Emily {\"O}hman}.} \bibinfo{year}{2020}\natexlab{}.
\newblock \showarticletitle{Challenges in Annotation: Annotator Experiences from a Crowdsourced Emotion Annotation Task.}. In \bibinfo{booktitle}{\emph{DHN}}. \bibinfo{pages}{293--301}.
\newblock


\bibitem[Penquitt et~al\mbox{.}(2025)]%
        {penquitt2025label}
\bibfield{author}{\bibinfo{person}{Sarina Penquitt}, \bibinfo{person}{Jonathan Klees}, \bibinfo{person}{Rinor Cakaj}, \bibinfo{person}{Daniel Kondermann}, \bibinfo{person}{Matthias Rottmann}, {and} \bibinfo{person}{Lars Schmarje}.} \bibinfo{year}{2025}\natexlab{}.
\newblock \showarticletitle{From Label Error Detection to Correction: A Modular Framework and Benchmark for Object Detection Datasets}.
\newblock \bibinfo{journal}{\emph{arXiv preprint arXiv:2508.06556}} (\bibinfo{year}{2025}).
\newblock


\bibitem[Peters et~al\mbox{.}(2024)]%
        {peters2024generalizable}
\bibfield{author}{\bibinfo{person}{Heinrich Peters}, \bibinfo{person}{Alireza Hashemi}, {and} \bibinfo{person}{James Rae}.} \bibinfo{year}{2024}\natexlab{}.
\newblock \showarticletitle{Generalizable Error Modeling for Human Data Annotation: Evidence From an Industry-Scale Search Data Annotation Program}.
\newblock \bibinfo{journal}{\emph{ACM Journal of Data and Information Quality}} \bibinfo{volume}{16}, \bibinfo{number}{3} (\bibinfo{year}{2024}), \bibinfo{pages}{1--15}.
\newblock


\bibitem[Pipino et~al\mbox{.}(2002)]%
        {pipino2002data}
\bibfield{author}{\bibinfo{person}{Leo~L. Pipino}, \bibinfo{person}{Yang~W. Lee}, {and} \bibinfo{person}{Richard~Y. Wang}.} \bibinfo{year}{2002}\natexlab{}.
\newblock \showarticletitle{Data quality assessment}.
\newblock \bibinfo{journal}{\emph{Commun. ACM}} \bibinfo{volume}{45}, \bibinfo{number}{4} (\bibinfo{year}{2002}), \bibinfo{pages}{211--218}.
\newblock
\href{https://doi.org/10.1145/505248.506010}{doi:\nolinkurl{10.1145/505248.506010}}


\bibitem[Ralph et~al\mbox{.}(2020)]%
        {ralph2020empirical}
\bibfield{author}{\bibinfo{person}{Paul Ralph}, \bibinfo{person}{Nauman bin Ali}, \bibinfo{person}{Sebastian Baltes}, \bibinfo{person}{Domenico Bianculli}, \bibinfo{person}{Jessica Diaz}, \bibinfo{person}{Yvonne Dittrich}, \bibinfo{person}{Neil Ernst}, \bibinfo{person}{Michael Felderer}, \bibinfo{person}{Robert Feldt}, \bibinfo{person}{Antonio Filieri}, \bibinfo{person}{Breno Bernard~Nicolau de França}, \bibinfo{person}{Carlo~Alberto Furia}, \bibinfo{person}{Greg Gay}, \bibinfo{person}{Nicolas Gold}, \bibinfo{person}{Daniel Graziotin}, \bibinfo{person}{Pinjia He}, \bibinfo{person}{Rashina Hoda}, \bibinfo{person}{Natalia Juristo}, \bibinfo{person}{Barbara Kitchenham}, \bibinfo{person}{Valentina Lenarduzzi}, \bibinfo{person}{Jorge Martínez}, \bibinfo{person}{Jorge Melegati}, \bibinfo{person}{Daniel Mendez}, \bibinfo{person}{Tim Menzies}, \bibinfo{person}{Jefferson Molleri}, \bibinfo{person}{Dietmar Pfahl}, \bibinfo{person}{Romain Robbes}, \bibinfo{person}{Daniel Russo}, \bibinfo{person}{Nyyti Saarimäki},
  \bibinfo{person}{Federica Sarro}, \bibinfo{person}{Davide Taibi}, \bibinfo{person}{Janet Siegmund}, \bibinfo{person}{Diomidis Spinellis}, \bibinfo{person}{Miroslaw Staron}, \bibinfo{person}{Klaas Stol}, \bibinfo{person}{Margaret-Anne Storey}, \bibinfo{person}{Davide Taibi}, \bibinfo{person}{Damian Tamburri}, \bibinfo{person}{Marco Torchiano}, \bibinfo{person}{Christoph Treude}, \bibinfo{person}{Burak Turhan}, \bibinfo{person}{Xiaofeng Wang}, {and} \bibinfo{person}{Sira Vegas}.} \bibinfo{year}{2020}\natexlab{}.
\newblock \bibinfo{title}{Empirical Standards for Software Engineering Research}.
\newblock
\showeprint[arxiv]{2010.03525}~[cs.SE]
\urldef\tempurl%
\url{https://arxiv.org/abs/2010.03525}
\showURL{%
\tempurl}


\bibitem[Ruggeri et~al\mbox{.}(2024)]%
        {ruggeri2024let}
\bibfield{author}{\bibinfo{person}{Federico Ruggeri}, \bibinfo{person}{Eleonora Misino}, \bibinfo{person}{Arianna Muti}, \bibinfo{person}{Katerina Korre}, \bibinfo{person}{Paolo Torroni}, {and} \bibinfo{person}{Alberto Barr{\'o}n-Cede{\~n}o}.} \bibinfo{year}{2024}\natexlab{}.
\newblock \showarticletitle{Let Guidelines Guide You: A Prescriptive Guideline-Centered Data Annotation Methodology}.
\newblock \bibinfo{journal}{\emph{arXiv preprint arXiv:2406.14099}} (\bibinfo{year}{2024}).
\newblock


\bibitem[Runeson and H{\"o}st(2009)]%
        {runeson2009guidelines}
\bibfield{author}{\bibinfo{person}{Per Runeson} {and} \bibinfo{person}{Martin H{\"o}st}.} \bibinfo{year}{2009}\natexlab{}.
\newblock \showarticletitle{Guidelines for conducting and reporting case study research in software engineering}.
\newblock \bibinfo{journal}{\emph{Empirical software engineering}}  \bibinfo{volume}{14} (\bibinfo{year}{2009}), \bibinfo{pages}{131--164}.
\newblock


\bibitem[Samuktha et~al\mbox{.}(2024)]%
        {samuktha2024framework}
\bibfield{author}{\bibinfo{person}{V Samuktha}, \bibinfo{person}{S Abhilash}, \bibinfo{person}{Nitish Kumar}, {and} \bibinfo{person}{P Rajalakshmi}.} \bibinfo{year}{2024}\natexlab{}.
\newblock \showarticletitle{A Framework for Object Classification via Camera-Radar Fusion with Automated Labeling}. In \bibinfo{booktitle}{\emph{2024 IEEE Sensors Applications Symposium (SAS)}}. IEEE, \bibinfo{pages}{1--6}.
\newblock


\bibitem[Schubert et~al\mbox{.}(2024)]%
        {schubert2024identifying}
\bibfield{author}{\bibinfo{person}{Marius Schubert}, \bibinfo{person}{Tobias Riedlinger}, \bibinfo{person}{Karsten Kahl}, \bibinfo{person}{Daniel Kr{\"o}ll}, \bibinfo{person}{Sebastian Schoenen}, \bibinfo{person}{Sini{\v{s}}a {\v{S}}egvi{\'c}}, {and} \bibinfo{person}{Matthias Rottmann}.} \bibinfo{year}{2024}\natexlab{}.
\newblock \showarticletitle{Identifying label errors in object detection datasets by loss inspection}. In \bibinfo{booktitle}{\emph{Proceedings of the IEEE/CVF Winter Conference on Applications of Computer Vision}}. \bibinfo{pages}{4582--4591}.
\newblock


\bibitem[Sharma(2020)]%
        {sharma2020evaluation}
\bibfield{author}{\bibinfo{person}{Devendra Sharma}.} \bibinfo{year}{2020}\natexlab{}.
\newblock \bibinfo{title}{Evaluation and Analysis of Perception Systems for Autonomous Driving}.
\newblock


\bibitem[Shen and Iandola(2020)]%
        {shen2020automated}
\bibfield{author}{\bibinfo{person}{Anting Shen} {and} \bibinfo{person}{Forrest~Nelson Iandola}.} \bibinfo{year}{2020}\natexlab{}.
\newblock \bibinfo{title}{Automated annotation techniques}.
\newblock
\newblock
\shownote{US Patent 10,872,251}.


\bibitem[van Dijck et~al\mbox{.}(2024)]%
        {van2024deciphering}
\bibfield{author}{\bibinfo{person}{Gijs van Dijck}, \bibinfo{person}{Carlos Aguilera}, {and} \bibinfo{person}{Shashank~M Chakravarthy}.} \bibinfo{year}{2024}\natexlab{}.
\newblock \showarticletitle{Deciphering disagreement in the annotation of EU legislation}.
\newblock \bibinfo{journal}{\emph{Artificial Intelligence and Law}} (\bibinfo{year}{2024}), \bibinfo{pages}{1--36}.
\newblock


\bibitem[Wac et~al\mbox{.}(2023)]%
        {wac2023capturing}
\bibfield{author}{\bibinfo{person}{Marceli Wac}, \bibinfo{person}{Raul Santos-Rodriguez}, \bibinfo{person}{Chris McWilliams}, {and} \bibinfo{person}{Christopher Bourdeaux}.} \bibinfo{year}{2023}\natexlab{}.
\newblock \showarticletitle{Capturing requirements for a data annotation tool for intensive care: Experimental user-centered design study}.
\newblock \bibinfo{journal}{\emph{arXiv preprint arXiv:2309.16500}} (\bibinfo{year}{2023}).
\newblock


\bibitem[Wand and Wang(1996)]%
        {wand1996anchoring}
\bibfield{author}{\bibinfo{person}{Yair Wand} {and} \bibinfo{person}{Richard~Y. Wang}.} \bibinfo{year}{1996}\natexlab{}.
\newblock \showarticletitle{Anchoring data quality dimensions in ontological foundations}.
\newblock \bibinfo{journal}{\emph{Commun. ACM}} \bibinfo{volume}{39}, \bibinfo{number}{11} (\bibinfo{year}{1996}), \bibinfo{pages}{86--95}.
\newblock
\href{https://doi.org/10.1145/240455.240479}{doi:\nolinkurl{10.1145/240455.240479}}


\bibitem[Wang et~al\mbox{.}(2024)]%
        {wang2024survey}
\bibfield{author}{\bibinfo{person}{Yuning Wang}, \bibinfo{person}{Zeyu Han}, \bibinfo{person}{Yining Xing}, \bibinfo{person}{Shaobing Xu}, {and} \bibinfo{person}{Jianqiang Wang}.} \bibinfo{year}{2024}\natexlab{}.
\newblock \showarticletitle{A survey on datasets for the decision making of autonomous vehicles}.
\newblock \bibinfo{journal}{\emph{IEEE Intelligent Transportation Systems Magazine}} (\bibinfo{year}{2024}).
\newblock


\bibitem[Watson et~al\mbox{.}(2023)]%
        {watson2023augmented}
\bibfield{author}{\bibinfo{person}{Eleanor Watson}, \bibinfo{person}{Thiago Viana}, {and} \bibinfo{person}{Shujun Zhang}.} \bibinfo{year}{2023}\natexlab{}.
\newblock \showarticletitle{Augmented Behavioral Annotation Tools, with Application to Multimodal Datasets and Models: A Systematic Review}.
\newblock \bibinfo{journal}{\emph{AI}} \bibinfo{volume}{4}, \bibinfo{number}{1} (\bibinfo{year}{2023}), \bibinfo{pages}{128--171}.
\newblock


\bibitem[Yang et~al\mbox{.}(2023)]%
        {yang2023uncertainties}
\bibfield{author}{\bibinfo{person}{Kai Yang}, \bibinfo{person}{Xiaolin Tang}, \bibinfo{person}{Jun Li}, \bibinfo{person}{Hong Wang}, \bibinfo{person}{Guichuan Zhong}, \bibinfo{person}{Jiaxin Chen}, {and} \bibinfo{person}{Dongpu Cao}.} \bibinfo{year}{2023}\natexlab{}.
\newblock \showarticletitle{Uncertainties in onboard algorithms for autonomous vehicles: Challenges, mitigation, and perspectives}.
\newblock \bibinfo{journal}{\emph{IEEE Transactions on Intelligent Transportation Systems}} (\bibinfo{year}{2023}).
\newblock


\bibitem[Zhong et~al\mbox{.}(2022)]%
        {zhong2022detecting}
\bibfield{author}{\bibinfo{person}{Ziyuan Zhong}, \bibinfo{person}{Zhisheng Hu}, \bibinfo{person}{Shengjian Guo}, \bibinfo{person}{Xinyang Zhang}, \bibinfo{person}{Zhenyu Zhong}, {and} \bibinfo{person}{Baishakhi Ray}.} \bibinfo{year}{2022}\natexlab{}.
\newblock \showarticletitle{Detecting multi-sensor fusion errors in advanced driver-assistance systems}. In \bibinfo{booktitle}{\emph{Proceedings of the 31st ACM SIGSOFT International Symposium on Software Testing and Analysis}}. \bibinfo{pages}{493--505}.
\newblock


\end{thebibliography}

\end{document}